\documentclass[
 reprint,
 superscriptaddress,
 amsmath,
 amssymb,
 aps,
 prb,
 citeautoscript,
 floatfix,
]{revtex4-2}

\usepackage{graphicx}
\usepackage{dcolumn}
\usepackage{bm}
\usepackage[dvipsnames]{xcolor}
\usepackage[colorlinks=true, citecolor={blue!80!black}, urlcolor={blue!50!black}, linkcolor = {blue!80!black}]{hyperref}
\usepackage{lipsum}  
\usepackage{upgreek}
\usepackage{siunitx}
\usepackage{mathtools}
\usepackage{verbatim}
\usepackage{xfrac}

\usepackage[toc]{appendix}
\usepackage{nameref,zref-xr}
\zxrsetup{toltxlabel}
\newcommand{\ket}[1]{\left|#1\right\rangle}

\begin{document}
% \myexternaldocument{supplement}

\preprint{APS/123-QED}

\title{Improved fluxonium readout through dynamic flux pulsing}

\author{Taryn V. Stefanski}
\thanks{Present address: QphoX, 2628 XG, Delft, Netherlands.}
\affiliation{QuTech and Kavli Institute of Nanoscience, Delft University of Technology, 2628 CJ, Delft, The Netherlands}
\affiliation{Quantum Engineering Centre for Doctoral Training, H. H. Wills Physics Laboratory and Department of Electrical and Electronic Engineering, University of Bristol, BS8 1FD, Bristol, UK}

\author{Figen Yilmaz}
\affiliation{QuTech and Kavli Institute of Nanoscience, Delft University of Technology, 2628 CJ, Delft, The Netherlands}

\author{Eugene Y. Huang}
\affiliation{QuTech and Kavli Institute of Nanoscience, Delft University of Technology, 2628 CJ, Delft, The Netherlands}

\author{Martijn F.S. Zwanenburg}
\affiliation{QuTech and Kavli Institute of Nanoscience, Delft University of Technology, 2628 CJ, Delft, The Netherlands}

\author{Siddharth Singh}
\affiliation{QuTech and Kavli Institute of Nanoscience, Delft University of Technology, 2628 CJ, Delft, The Netherlands}

\author{Siyu Wang}
\affiliation{QuTech and Kavli Institute of Nanoscience, Delft University of Technology, 2628 CJ, Delft, The Netherlands}

\author{Lukas J. Splitthoff}
\affiliation{QuTech and Kavli Institute of Nanoscience, Delft University of Technology, 2628 CJ, Delft, The Netherlands}

\author{Christian Kraglund Andersen}
\affiliation{QuTech and Kavli Institute of Nanoscience, Delft University of Technology, 2628 CJ, Delft, The Netherlands}

\date{\today}

\begin{abstract}
The ability to perform rapid, high fidelity readout of a qubit state is an important requirement for quantum algorithms and, in particular, for enabling operations such as mid-circuit measurements and measurement-based feedback for error correction schemes on large quantum processors. The growing interest in fluxonium qubits, due to their long coherence times and high anharmonicity, merits further attention to reducing the readout duration and measurement errors. We find that this can be accomplished by exploiting the flux tunability of fluxonium qubits. In this work, we experimentally demonstrate flux-pulse-assisted readout, as proposed in \href{https://doi.org/10.1103/PhysRevApplied.22.014079}{Phys. Rev. Applied \textbf{22}, 014079}, in a setup without a quantum-limited parametric amplifier. Increasing the dispersive shift magnitude by almost 20\% through flux pulsing, we achieve an assignment fidelity of 94.3\% with an integration time of 280 ns. The readout performance is limited by state initialization, but we find that the limit imposed only by the signal-to-noise ratio corresponds to an assignment fidelity of 99.9\% with a 360 ns integration time. We also verify these results through simple semi-classical simulations. These results constitute the fastest reported readout of a fluxonium qubit, with the prospect of further improvement by incorporation of a parametric amplifier in the readout chain to enhance measurement efficiency.
\end{abstract}

\maketitle
\section{Introduction}
Superconducting qubit-based architectures offer a promising route towards enabling quantum computation~\cite{kjaergaard2020superconducting}. The ability to engineer these circuits by tailoring the relative energy parameters of the constituent components has resulted in many qubit variations~\cite{nakamura1999coherent, koch2007charge, manucharyan2009fluxonium, martinis2002rabi, orlando1999superconducting, brooks2013protected}. Of these different qubit platforms, the transmon qubit has often been the preferred choice for scaling these devices~\cite{krinner2022realizing, marques2022logical, acharya2024quantum, kim2023evidence}. However, the transmon qubit suffers from a low anharmonicity and limited relaxation times due, primarily, to dielectric loss~\cite{eun2023shape, wang2015surface}.

With the aim of overcoming these limitations, the fluxonium qubit has recently gained increasing attention as a potential alternative. The fluxonium, composed of a shunting inductance and capacitance in parallel with a Josephson junction, has a typical frequency of $\omega_{q}/2\pi \lesssim$ 1 GHz and an anharmonicity of a few GHz at its standard operation point, referred to as the sweet-spot, of $\Phi_{\mathrm{ext}}/\Phi_{0} = 0.5$, where $\Phi_{\mathrm{ext}}$ is the flux threading the superconducting loop and $\Phi_{0}=h/(2e)$ is the magnetic flux quantum~\cite{nguyen2019high}. The combination of the low frequency of the fluxonium qubit with its small charge matrix element has enabled state-of-the-art coherence times exceeding 1 ms~\cite{ding2023high, somoroff2023millisecond}. Furthermore, the large anharmonicity lends itself favorably to high fidelity single qubit gates, which have reached fidelities exceeding 99.99\%~\cite{rower2024suppressing}. High fidelity two qubit gates between fluxonium qubits have additionally been demonstrated with both direct capacitive coupling, and mediated by a tunable coupler~\cite{moskalenko2022high, dogan2022demonstration, ding2023high}.

In order to prove advantageous as the building block for future quantum devices, the speed and fidelity of initialization and readout of these qubits must also be considered. While various methods for achieving high fidelity state preparation have been reported, optimization of fluxonium readout remains an outstanding challenge~\cite{zhang2021universal, gusenkova2021quantum, gebauer2020state, wang2024efficient}. In this work, we demonstrate fast readout of a fluxonium qubit without using a parametric amplifier, achieving an assignment fidelity of 94.3\% in 280 ns and an assignment fidelity of 99.9\% in 360 ns if we only consider the limit imposed by the signal-to-noise ratio (SNR), which we will refer to as the SNR-limited fidelity. These results are accomplished through fast flux tuning of the qubit in order to enhance the dispersive interaction strength, as proposed for fluxoniums in Ref.~\cite{stefanski2024flux}, and similarly demonstrated with transmons in Ref.~\cite{swiadek2023enhancing}. Our experiments were performed without the use of quantum-limited parametric amplifiers, thus, future experiments could decrease the readout duration further~\cite{white2023readout, macklin2015near, frattini2018optimizing, vijay2011observation, yamamoto2008flux}. Nonetheless, we show substantial speed-up compared to other instances of fluxonium readout, typically utilizing integration times on the order of a microsecond, while also achieving comparable performance. These results support the feasibility of fast fluxonium readout and have the potential to be further improved with modifications to the readout chain.

In this paper, we first introduce our device under test and its initial characterization with spectroscopic measurements in order to identify a suitable readout point. The subsequent section compares the readout performance with flux-pulse-assisted readout versus standard readout at the sweet-spot. We define the quantities used to compare the results and describe how we numerically model both cases, for which we find good agreement with the measured data. Finally, we comment on the implications of our findings and suggest avenues for improvement.

\section{Fluxonium Device}
\begin{figure}
\centering
\includegraphics[scale=0.47]{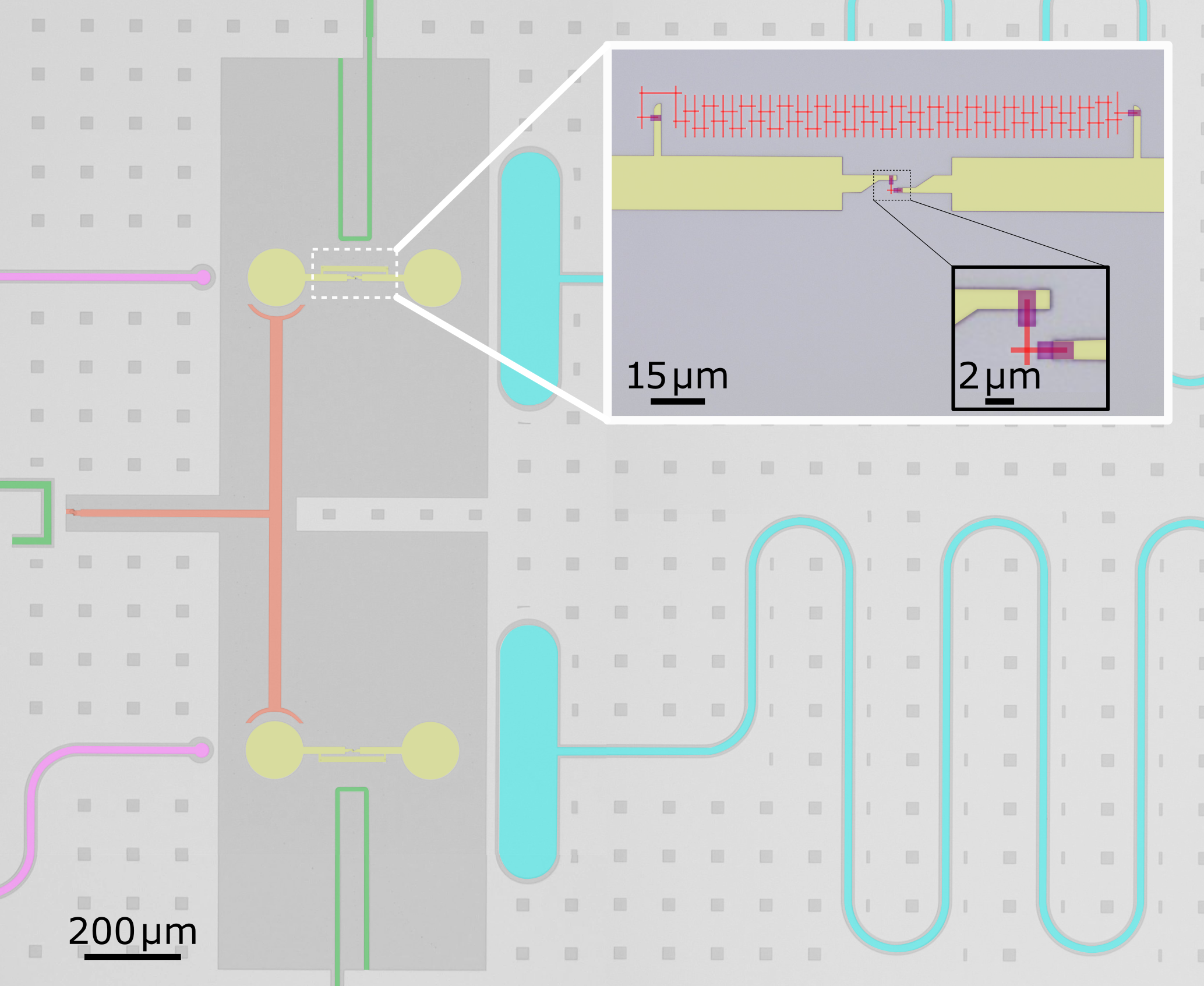}
\caption{False-colored optical image of a pair of fluxonium qubits (yellow) coupled via a tunable transmon qubit (orange) from our device under test. We address a single fluxonium within the set to demonstrate flux-pulse-assisted readout. Each qubit is inductively coupled to a flux line (green) and each fluxonium is capacitively coupled to its own charge line (pink) and readout resonator (blue). The inset magnifies the fluxonium flux loop defined by an inductor composed of an array of 100 Josephson junctions (top) and a single Josephson junction (bottom), which is further magnified. The junction electrodes (red) are galvanically connected to the fluxonium capacitor pads with patches (purple).}
\label{fig:device}
\end{figure}

Our fluxonium qubit, see Fig.~\ref{fig:device}, is part of a three-qubit system in which two differential fluxonium qubits are capacitively coupled by a grounded, tunable transmon qubit, where fabrication details are provided in App.~\ref{app:fab}. In this work, we address a single fluxonium qubit which we flux bias at $\Phi_{\mathrm{ext}} = 0.5\Phi_0$. Additionally, we bias both the transmon and other fluxonium at their respective zero external flux bias points. We maintain this condition by compensating for residual DC flux crosstalk through application of the inverted crosstalk matrix, assuming a linear combination of the contributions from each of the qubits~\cite{abrams2019methods}. As a result, the remainder of this work pertains only to the individual fluxonium qubit on which the readout scheme was implemented.

\begin{figure}
\centering
\includegraphics[scale=1]{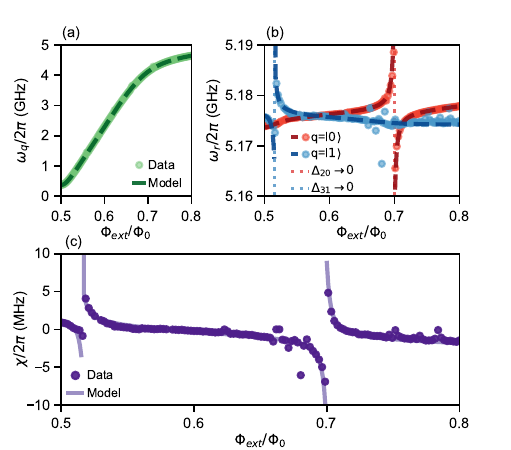}
\caption{(a) Qubit spectroscopy as a function of external flux bias. (b) Measured (points) and fitted (dashed curves) resonator spectroscopy as a function of external flux bias for the qubit prepared in the ground (red) or excited (blue) state. Vertical asymptotes denote where the $\ket{2}\rightarrow\ket{0}$ (red dotted) and $\ket{3}\rightarrow\ket{1}$ (blue dotted) fluxonium transition frequencies equal the resonator frequency. (c) Measured dispersive shift (dark points) as a function of external flux bias with the numerical prediction indicated by the solid curve.}
\label{fig:spec and dis shift}
\end{figure}

We measure the qubit frequency, $\omega_{q}$, and resonator frequency, $\omega_{r}$, for the qubit prepared in either $\ket{0}$ or $\ket{1}$ as a function of external flux bias with standard spectroscopy measurements, see Fig.~\ref{fig:spec and dis shift}(a) and (b). As a result of the low frequency of the fluxonium such that $\hbar\omega_{q}\lesssim k_{\mathrm{B}}T$ near the sweet-spot, there exists a non-negligible excited state population due to thermal excitations. To measure the state-dependent resonator frequency when $\omega_{q}/2\pi \lesssim 1$ GHz, we post-select the spectroscopy data by first applying an $X_{\pi/2}$ rotation to the qubit to prepare it in an equal superposition, after which we measure the qubit to collapse it onto an eigenstate of $\hat{\sigma}_{z}$, before ultimately applying the resonator spectroscopy tone. For $\omega_{q}/2\pi > 1$ GHz, we idle the qubit in order to allow any residual excitations to relax to the ground state, and subsequently apply the spectroscopy pulse in order to measure $\omega_{r,q=\ket{0}}$. We repeat this for measurements of $\omega_{r,q=\ket{1}}$ with the addition of an $X_{\pi}$ rotation prior to the spectroscopy pulse in order to prepare the qubit in its excited state. We additionally use the measured resonator frequencies to extract the dispersive shift as a function of external flux according to $\chi = \left(\omega_{r,q=\ket{1}}-\omega_{r,q=\ket{0}}\right)/2$, see Fig.~\ref{fig:spec and dis shift}(c).

From the measured spectroscopy data, we extract fluxonium energy parameters of $E_{J}/2\pi = 3.82$ GHz, $E_{C}/2\pi = 0.865$ GHz, and $E_{L}/2\pi = 0.822$ GHz, as well as a bare resonator frequency of $\omega_{r}/2\pi = 5.175$ GHz and a qubit-resonator coupling strength of $g/2\pi = 37.2$ MHz through modeling of the coupled qubit-resonator system with \texttt{scQubits}~\cite{chitta2022computer, groszkowski2021scqubits}. Given these quantities, we simulate the dispersive shift following Ref.~\cite{stefanski2024flux}, which demonstrates excellent agreement with the measured values. At sweet-spot, the fluxonium has a frequency of $\omega_{q}/2\pi = 377$ MHz and is dispersively coupled to the readout resonator with state-dependent frequencies of $\omega_{r,q=\ket{0}}/2\pi = 5.1739$ GHz and $\omega_{r,q=\ket{1}}/2\pi = 5.1757$ GHz, resulting in a dispersive interaction strength of $\chi/2\pi = 0.92$ MHz. We extract a resonator linewidth of $\kappa/2\pi = 6.04$ MHz by fitting the resonator transmission data at $\Phi_{\mathrm{ext}}/\Phi_{0} = 0.5$.

In an effort to locate a suitable readout point, we additionally identify the origins of the diverging features in the resonator spectra through examination of the higher-order transition frequencies of the fluxonium qubit. As the detuning, $\Delta_{ij}$, between a qubit transition $\ket{i}\rightarrow\ket{j}$ and the resonator frequency tends to zero, the corresponding contribution to the total dispersive shift diverges~\cite{zhu2013circuit}. As a result of including these higher-order terms in our model, we identify the $\ket{3}\rightarrow\ket{1}$ $\left(\ket{2}\rightarrow\ket{0}\right)$ transition as the cause of the feature at $\Phi_{\mathrm{ext}}/\Phi_{0} \approx 0.517~(0.70)$. The larger level repulsion around $\Phi_{\mathrm{ext}}/\Phi_{0} = 0.70$, compared to that near $\Phi_{\mathrm{ext}}/\Phi_{0} = 0.517$, enables measurement of large dispersive shift values, up to $\vert\chi\vert/2\pi = 7$ MHz, over a wider range of external flux bias points. This observation motivates our choice to exploit this feature in order to enhance readout performance through accessing an increased dispersive shift magnitude.

The low frequency of the fluxonium additionally necessitates active reset prior to measurement. Our reset is accomplished by deploying a series of successive net-zero flux pulses. This pulse sequence temporarily shifts the fluxonium away from sweet-spot such that it has a higher frequency and residual excited state population relaxes to the ground state~\cite{moskalenko2022high}. The choice of net-zero flux pulses, with alternating positive and negative amplitudes with equal magnitudes, prevents the accumulation of distortions in the flux bias line on a long time scale~\cite{rol2019fast, zhang2021universal}. We use this initialization scheme prior to all measurements presented in the remainder of this text. 

\section{Flux-Pulse-Assisted Readout}
In tuning up a target readout point for flux-pulse-assisted readout, we measure the state assignment error as a function of readout and flux pulse amplitudes with a fixed integration time of 200 ns and flux pulse rise time of 50 ns. We observe the highest fidelity using $\Bar{n} \approx 52$ readout resonator photons and shifting the qubit to $\Phi_{\mathrm{ext}}/\Phi_{0} = 0.6567$, see App.~\ref{app:cal} for details regarding the calibration. At this flux bias point, the qubit frequency is $\omega_{q}/2\pi = 3.47$ GHz, the dispersive shift is $\chi/2\pi = -1.09$ MHz, and we measure a relaxation time of approximately $T_{1} = 10~\mu$s. Due to the absence of a parametric amplifier, we are limited to a measurement efficiency of $\eta \approx 6\%$, see App.~\ref{app:cal}, which in turn necessitates the large photon number for readout. We suspect that we are unable to pulse closer to the divergence in the dispersive shift as the large number of photons breaks the dispersive approximation for the higher-lying states, leading to measurement-induced state transitions~\cite{blais2004cavity, blais2021circuit, gambetta2006qubit, khezri2023measurement, sank2016measurement}. We may be further limited by the fact that the $\ket{6}\rightarrow\ket{0}$ fluxonium transition approaches a multiple of the resonator frequency, i.e. $\omega_{60} = 3\omega_{r}$, around $\Phi_{\mathrm{ext}}/\Phi_{0} = 0.64$, which could lead to measurement-induced state transitions, especially given the large photon number necessitated by our setup~\cite{nesterov2024measurement}. However, we find that this modest increase in dispersive shift accessed through flux pulsing is sufficient to improve our readout fidelity.

\begin{figure}
\centering
\includegraphics[scale=1]{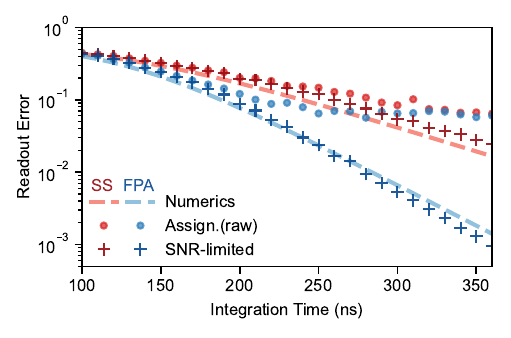}
\caption{Assignment (circles) and SNR-limited (crosses) readout error as a function of integration time for readout at sweet-spot (SS, red) and employing flux-pulse-assisted readout (FPA, blue) with a flux bias shift of $\Delta\Phi_{\mathrm{ext}}/\Phi_{0} = 0.1567$. Dashed curves are the result of numerical modeling.}
\label{fig:readout}
\end{figure}

To compare the performance of conventional readout at the sweet-spot to flux-pulse-assisted readout, we measure the assignment and SNR-limited error as a function of integration time for each method, see Fig.~\ref{fig:readout}. The assignment error is given by 
\begin{equation}
    E_{\mathrm{assign.}} = \frac{P(0|1)+P(1|0)}{2},
    \label{assignment fidelity}
\end{equation}
where $P(x|y)$ denotes the probability of measuring $\ket{x}$ for a state prepared in $\ket{y}$. This metric captures the effect of errors leading to incorrect state assignment including imperfect initialization, relaxation during measurement, and measurement-induced state transitions. Alternatively, the SNR-limited error considers only the fitted distributions of correctly assigned shots. Fitting the $\ket{0}$ and $\ket{1}$ state distributions each with a single Gaussian curve, the SNR is given by
\begin{equation}
    \mathrm{SNR} = \frac{|\mu_{\ket{1}}-\mu_{\ket{0}}|}{\sqrt{\sigma_{\ket{0}}^{2}+\sigma_{\ket{1}}^{2}}},
    \label{SNR}
\end{equation}
where $\mu_{\ket{i}}$ and $\sigma_{\ket{i}}$ are the mean and standard deviations, respectively, for the state $\ket{i}$. The resulting SNR-limited error follows as
\begin{equation}
    E_{\mathrm{SNR-lim.}} = \frac{1}{2}\mathrm{erfc}\left(\frac{\mathrm{SNR}}{2}\right),
    \label{separation error}
\end{equation}
\noindent which quantifies the separation between the two distributions.

We perform heterodyne measurement of the readout signal, applied at a frequency of $\omega_{\mathrm{RO}}/2\pi = 5.1747$ GHz, where the exact value is within 100 kHz of $\omega_{r,q=\ket{0},\mathrm{SS}}+\chi_{\mathrm{SS}}$, and the subscript `SS' refers to the value measured at the sweet-spot. At the flux-pulse-assisted (FPA) readout point, this readout frequency is detuned from $\left(\omega_{r,q=\ket{0},\mathrm{FPA}}+\omega_{r,q=\ket{1},\mathrm{FPA}}\right)/2$, and is instead closer to $\omega_{r,q=\ket{1},\mathrm{FPA}}$, i.e. $\vert\omega_{\mathrm{RO}}-\omega_{r,q=\ket{1},\mathrm{FPA}}\vert\approx 2\pi\times380$ kHz, compared to $\vert\omega_{\mathrm{RO}}-\omega_{r,q=\ket{0},\mathrm{FPA}}\vert\approx2\pi\times2.56$ MHz. To account for this deviation when modeling the expected output field, we modify the differential equation of the coherent state amplitude of the intracavity field in Ref.~\cite{stefanski2024flux} such that it reads
\begin{equation}
    \dot{\alpha} = -i\Delta_{\pm}\left\langle\hat{\sigma}_{z}\right\rangle\alpha - \frac{1}{2}\kappa\alpha - \sqrt{\kappa}\alpha_{\mathrm{in}},
    \label{langevinalpha}
\end{equation}
\noindent where we maintain the boundary condition $\alpha_{\mathrm{out}} = \alpha_{\mathrm{in}} + \sqrt{\kappa}\alpha$~\cite{didier2015fast, didier2015heisenberg, gardiner2004quantum}. In Eq.~\ref{langevinalpha}, we have 
\begin{equation}
    \Delta_{\pm} = \left(\frac{\omega_{r, q=\ket{0}}(t)+\omega_{r, q=\ket{1}}(t)}{2}-\omega_{\mathrm{RO}}\right)\pm\chi(t),
\end{equation}
\noindent where $+$($-$) corresponds to the $\hat{\sigma}_{z}$ eigenvalue of $+1$($-1$) associated with the $\ket{1}$($\ket{0}$) state~\cite{blais2021circuit}. Similarly, assuming our coherent input field can be expressed as $\alpha_{\mathrm{in}} = -\epsilon/\sqrt{\kappa}$, $\epsilon$ becomes
\begin{equation}
    \epsilon = \sqrt{\frac{2\Bar{n}}{\left(\Delta_{+}^2+\frac{\kappa^{2}}{4}\right)^{-1} + \left(\Delta_{-}^2+\frac{\kappa^{2}}{4}\right)^{-1}}}.
\end{equation}
\noindent Numerically solving for the cavity output field as a function of integration time in order to obtain the SNR, and subsequently, the SNR-limited error, we obtain the theoretical curves in Fig.~\ref{fig:readout} using an average resonator photon number of $\bar{n} = 75$. This demonstrates excellent qualitative agreement with our data, thus, verifying the ability of the simple theoretical model to predict the experimental data. We attribute the difference in $\bar{n}$ from the experimentally extracted value to uncertainties in our measurement efficiency calibration data. 

Overall, we find that flux-pulse-assisted readout outperforms standard readout at sweet-spot, reaching 99.9\% SNR-limited readout fidelity in 360 ns, compared to 97.5\% in the latter case for the same integration time. This is accomplished with a mild increase in the dispersive shift magnitude by $\sim 20\%$. This result suggests that we may be able to further shorten the integration times or use a lower photon number for readout through enhancing our measurement efficiency. We can additionally compare the assignment error to the SNR-limited error to gain intuition about other mechanisms limiting our overall readout. While the flux-pulse-assisted assignment fidelity exceeds that at sweet-spot by at least 5\% for integration times between 140-280 ns, both plateau from about 300 ns with an error rate of about 6\%. This suggests that both become limited by relaxation and imperfect state preparation, which could be enhanced in the future with the use of flux-pulse-assisted sideband driving, as in Ref.~\cite{wang2024efficient}.

\begin{figure}[b]
\centering
\includegraphics[scale=1]{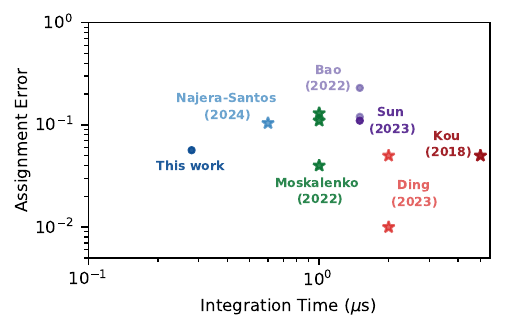}
\caption{Reported assignment error of fluxonium readout from Refs.~\cite{bao2022fluxonium, ding2023high, kou2018simultaneous, moskalenko2022high, najera2024high, sun2023characterization}, where stars indicate the use of a parametric amplifier as the first amplification stage in the readout chain. Not included are Ref.~\cite{zhang2021universal} with a reported 50\% assignment error using plasmon-assisted readout and Ref.~\cite{wang2024efficient} with average assignment errors of 16.6\% and 14.7\% for two fluxonium qubits, achieved with unspecified integration times. See App.~\ref{app:compare} and Table~\ref{tab:lit readout} for details.}
\label{fig:compare}
\end{figure}

\section{Conclusion}
In conclusion, we have demonstrated the utility of flux-pulse-assisted readout for fluxonium qubits, measuring a readout assignment fidelity of  94.3\% in 280 ns. Furthermore, we achieve a 99.9\% SNR-limited readout fidelity in 360 ns, compared to 97.5\% with conventional readout at the sweet-spot, and find that these SNR-limited fidelities are well predicted by a simple semi-classical simulation. Our results were accomplished by employing a flux pulse during readout in order to dynamically increase the dispersive shift by about 20\%. To the best of our knowledge, we have demonstrated the fastest readout of a fluxonium qubit, measuring assignment fidelity with a measurement efficiency of $\eta = 6\%$, see Fig.~\ref{fig:compare}. Our results are comparable with other values reported, also considering works in which a parametric amplifier was used at the first stage of amplification. We anticipate that the inclusion of a quantum-limited parametric amplifier in our measurement setup will allow for further improvement of the achievable assignment fidelity as we will require fewer photons for readout, which may allow us to flux to a point with a larger dispersive shift~\cite{eichler2014controlling, esposito2021perspective, macklin2015near}. We may consider additional refinements through advanced pulse shaping techniques and more rigorous distortion corrections at cryogenic temperatures~\cite{rol2020time}. Overall, this readout method enables the fast readout of fluxonium qubits that are well protected from thermal dephasing at sweet-spot through dynamic flux pulsing during readout.

\section*{Acknowledgements}
T.V.S. acknowledges the support of the Engineering and Physical Sciences Research Council (EPSRC) under EP/SO23607/1. This work was partly funded through the support of Holland High Tech (TKI) under project 00PPS334 and from the Dutch Research Council (NWO). 

T.V.S., E.Y.H., and M.F.S.Z. characterized the device and performed the experiment. T.V.S. analyzed the data and simulated the numerical model. F.Y. and S.W. designed the device. F.Y., S.S., and L.J.S. fabricated the device. C.K.A. supervised the project. T.V.S. wrote the manuscript with input from all authors.

\section*{Data Availability}
All of the measurement data acquired during this experiment has been made available~\cite{dataset}. The repository containing the code used to analyze the data and perform the numerical modeling has additionally been made available~\cite{code_repo}.

\appendix
\renewcommand{\thesection}{\Alph{section}}
\section{Fabrication of Device}
\label{app:fab}
Our 9$\times$9mm$^{2}$ device is fabricated on a 525~$\mu$m thick high-resistivity ($>$20k$\Omega$cm) silicon substrate from Topsil with 1-0-0 orientation. The base layer is a 200~nm thin film of NbTiN, sputtered using a Nb:Ti target with 70:30 ratio by weight and an N$_{2}$ flow rate of 5~SCCM. The circuit structures are patterned with electron-beam lithography (Raith EBPG5200) and subsequently developed with pentyl acetate for 60~s, followed by xylene for 5~s, and isopropanol (IPA) for 30~s. The development is followed by a two-step dry etching process (Sentech Etchlab 200): (i) SF$_{6}$:O$_{2}$ = 13.5:5 and (ii) SF$_{6}$:O$_{2}$ = 4:15. The resist is removed first with O$_{2}$ plasma (PVA Tepla 300), followed by submersion in PRS-3000 for 2 hours at 40\textdegree C, and N-Methylpyrrolidone (NMP) overnight. The sample is then cleaned with HNO$_{3}$ for 60~s and buffered oxide etch (BOE) (7:1) for 30 minutes, and rinsed in two baths of DI water for 10~s and 60~s, after which it is blow dried with N$_{2}$.

Prior to the junction deposition, we spin a bi-layer resist stack of MMA-EL8 and PMMA-950. The Al/AlOx/Al Manhattan-style Josephson junctions and undercuts are patterned with electron-beam lithography (Raith EBPG5200). For the junctions, we use a current of 10~nA with a dose of 1300~$\mu$C/cm$^{2}$, and for the undercuts we use a current of 16~nA with a dose of 540~$\mu$C/cm$^{2}$. Development is done with a 1:3 mixture of H$_{2}$O:IPA at 6\textdegree C for 90~s and IPA for an additional 30~s. We again clean the sample with O$_{2}$ plasma (PVA Tepla 300), followed by BOE (7:1) for 40~s, rinsed in two baths of DI water for 10~s and 60~s, and blow dried with N$_{2}$. We perform successive evaporation of junction electrodes (Plassys MEB 550) with 30~nm and 110~nm thicknesses, where we oxidize each electrode with a pressure of 1.3~mbar for 11 minutes. Lift-off is done with submersion in a 50\textdegree C acetone bath for 2 hours, followed by an 80\textdegree C NMP bath for 2 hours, with a final rinse in acetone and IPA for 1 minute each. We additionally deposit Al patches to ensure galvanic connection between the junction electrodes and capacitor pads. The procedure for the patch deposition and subsequent lift-off and cleaning follows that of the junctions, with the use of a 10~nA current and 1200~$\mu$C/cm$^{2}$ dose during the e-beam lithography, and the addition of ion milling prior to the aluminum evaporation to obtain a 150~nm thickness.

\section{Hardware Configuration}
\label{app:hardware config}
\begin{figure}[h]
\centering
\includegraphics[scale=0.8]{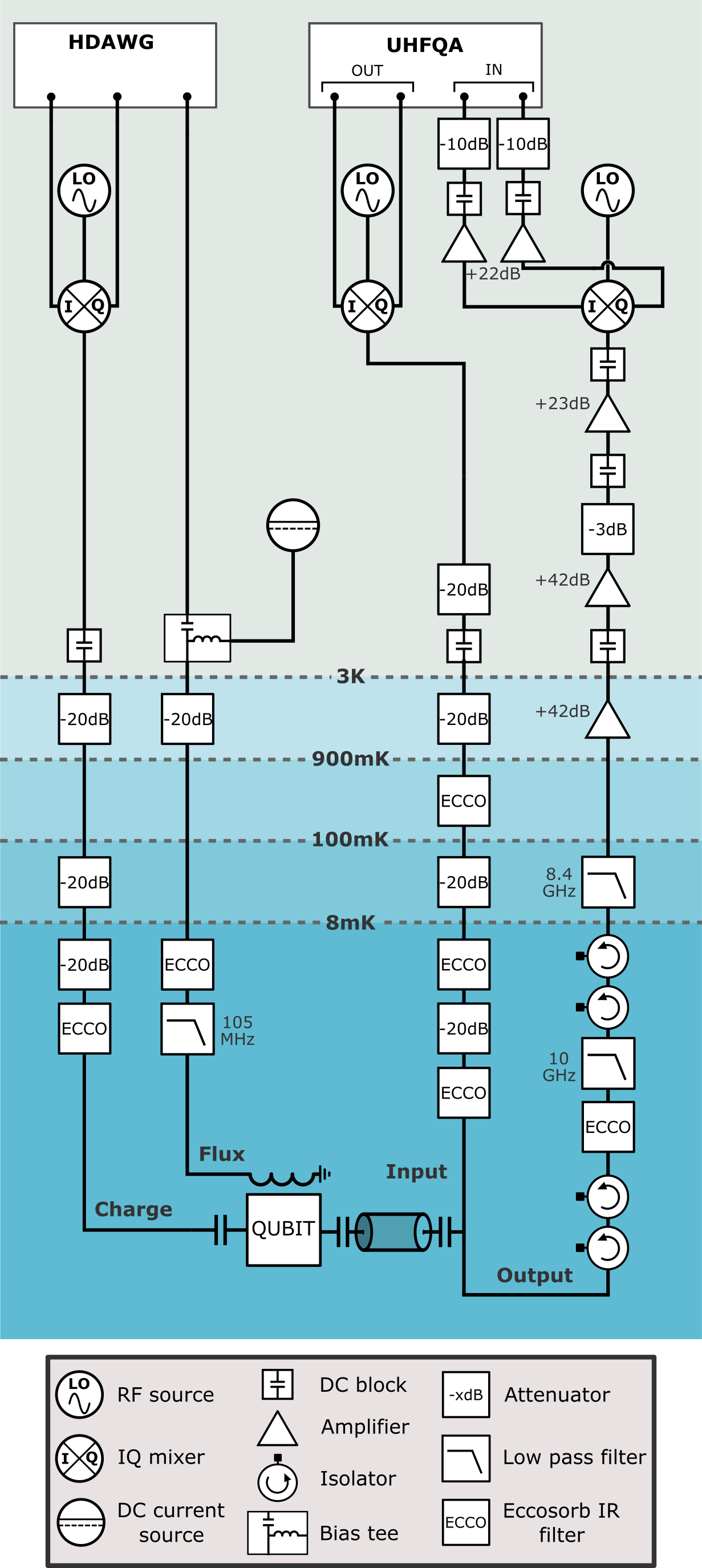}
\caption{Detailed schematic of the experimental setup.}
\label{fig:fridge}
\end{figure}

We install our device at the 8mK stage of a Bluefors LD400 dilution refrigerator, where it is encapsulated by an aluminum shield, and further thermalized and shielded from electromagnetic radiation by a copper can and two Mu-metal cans. In total, our device is equipped with 8 charge lines, 12 flux lines, one input line, and one output line, where the configuration of microwave components and room temperature electronics is detailed in Fig.~\ref{fig:fridge}.

The generation of input signals and processing of output signals is achieved with the Zurich Instruments UHFQA quantum analyzer. These signals are up or downconverted, respectively, using the Zurich Instruments HDIQ IQ modulator, with the local oscillator (LO) signal supplied by an APMS20G-4 AnaPico signal generator. For flux bias points at which the qubit frequency is $\omega_{q}/2\pi\leq 1$ GHz, we use control signals directly synthesized by the Zurich Instruments HDAWG 750 MHz 8-channel arbitrary waveform generator. Above $\omega_{q}/2\pi = 1$ GHz, we upconvert signals from the HDAWG using an IQ-1545LMP Marki Microwave mixer, with the LO provided by the AnaPico. The latter configuration is depicted in Fig.~\ref{fig:fridge}. Our flux pulses are additionally generated by the HDAWG, and constant current biases are provided by the S4g current source module of our SPI-Rack, which are combined into a common flux line with a bias-tee at room temperature~\cite{S4gDocumentation, SPIRackDocumentation}.

All lines are equipped with various microwave components including attenuators, Eccosorb filters, and low pass filters, where the associated attenuations and cut-off frequencies are listed in Fig.~\ref{fig:fridge}. The Eccosorb filters are machined and assembled in-house, and those on the charge, flux, and input lines are made with grade CR-124, and the Eccosorb filter on the output line is made with grade CR-110. On the output line, we additionally use two LNF-ISISC4\_8A cryogenic dual
junction isolators from Low Noise Factory at base temperature. The output signal is amplified by a Low Noise Factory LNF-LNC4\_8C HEMT amplifier at 4K. At room temperature, the signal is further amplified by a Low Noise Factory LNF-LNR4\_8ART amplifier, Mini-Circuits ZRON-8G+ amplifier, and Mini-Circuits GALI-3+ amplifier(s).

\section{Calibration}
\label{app:cal}
\begin{table*}[htbp]
\caption{Overview of reported fluxonium readout assignment fidelities, as illustrated in Fig.~\ref{fig:compare}, and fluxonium qubit parameters. For Refs.~\cite{bao2022fluxonium, moskalenko2022high, wang2024efficient}, where results were reported for two or three fluxoniums, the corresponding values are indexed with a lettered subscript matching the in-text naming convention used. The * symbol indicates the $\ket{0}\rightarrow\ket{1}$ transition dispersive shift value at the sweet-spot, which differs from the dispersive shift value exploited to obtain the readout reported. The reported assignment fidelity from Ref.~\cite{zhang2021universal} was achieved using the $\ket{g}\rightarrow\ket{f}$ plasmon transition dispersive shift with a value of $\chi/2\pi = 0.3$ MHz. The final column indicates the presence (or lack) of a parametric amplifier and specifies the implementation where applicable - traveling wave parametric amplifier (TWPA), Josephson parametric converter amplifier (JPCA), or impedance matched parametric amplifier (IMPA). Values not reported are listed as NR.}
\label{tab:lit readout}
\renewcommand{\arraystretch}{1.25}
\setlength{\tabcolsep}{11pt}
\begin{tabular}{lcccccc}
\hline \hline
 & \multicolumn{1}{l}{Assign. Fid.} & \multicolumn{1}{l}{Int. Time} & $\omega_{q}/2\pi$ & $\kappa/2\pi$ & $\chi_{SS}/2\pi$ & \multicolumn{1}{l}{} \\
Ref. & (\%) & ($\mu$s) & (MHz) & (MHz) & (MHz) & Para. amp.? \\ \hline
Bao (2022)~\cite{bao2022fluxonium} & \begin{tabular}[c]{@{}c@{}}88$_{A}$\\ 77$_{B}$\end{tabular} & 1.5 & \begin{tabular}[c]{@{}c@{}}1090$_{A}$\\ 1330$_{B}$\end{tabular} & 14.3 & 0.63 & No \\ \hline
Ding (2023)~\cite{ding2023high} & 95-99 & 2.0 & 242-426 & 1.5 & 0.3 & Yes (TWPA) \\ \hline
Kou (2018)~\cite{kou2018simultaneous} & 95 & 5.0 & \begin{tabular}[c]{@{}c@{}}565$_{A}$\\ 579$_{B}$\end{tabular} & \begin{tabular}[c]{@{}c@{}}10$_{A}$\\ 14$_{B}$\end{tabular} & NR & Yes (2xJPCAs) \\ \hline
Moskalenko (2022)~\cite{moskalenko2022high} & \begin{tabular}[c]{@{}c@{}}87$_{A}$\\ 89$_{B}$\\ 96$_{S}$\end{tabular} & 1.0 & \begin{tabular}[c]{@{}c@{}}688$_{A}$\\ 665$_{B}$\\ 750$_{S}$\end{tabular} & \begin{tabular}[c]{@{}c@{}}7.6$_{A}$\\ 7.2$_{B}$\\ 7.4$_{S}$\end{tabular} & \begin{tabular}[c]{@{}c@{}}0.26$_{A}$\\ 0.27$_{B}$\\ 0.5$_{S}$\end{tabular} & Yes (IMPA) \\ \hline
Najera-Santos (2024)~\cite{najera2024high} & 89.6 & 0.6 & 1.8 & 2.4 & NR & Yes (TWPA) \\ \hline
Sun (2023)~\cite{sun2023characterization} & 89 & 1.5 & 385 & NR & NR & No \\ \hline
Wang (2024)~\cite{wang2024efficient} & \begin{tabular}[c]{@{}c@{}}83.4$_{A}$\\ 85.3$_{B}$\end{tabular} & NR & \begin{tabular}[c]{@{}c@{}}696$_{A}$\\ 693$_{B}$\end{tabular} & \begin{tabular}[c]{@{}c@{}}25$_{A}$\\ 17$_{B}$\end{tabular} & NR & No \\ \hline
Zhang (2021)~\cite{zhang2021universal} & 50 & NR & 14 & NR & 0.06* & No \\ \hline
This work & 94.3 & 0.28 & 377 & 6.04 & 0.92* & No \\ \hline \hline
\end{tabular}
\end{table*}

\begin{figure}[h]
\centering
\includegraphics[scale=1]{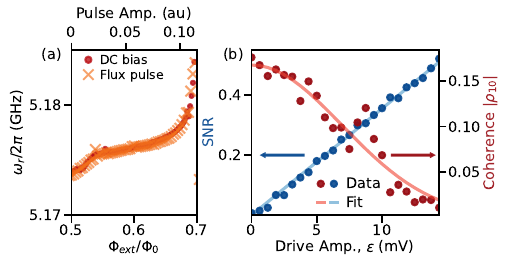}
\caption{(a) Resonator frequency when the qubit is prepared in $\ket{0}$ as a function of external flux bias, accessed via constant DC current bias (red circles) or through applying a flux pulse (orange $\times$'s). The aligned data points indicate the room temperature flux pulse amplitude (top x-axis) corresponding to a given external flux bias. (b) Signal to noise ratio (blue) and coherence (red) as a function of measurement pulse amplitude. Data (points) acquired with variable strength single shot readout and Ramsey measurements and fit with linear and Gaussian curves, respectively.}
\label{fig:cal}
\end{figure}

To calibrate the flux pulse amplitude for our flux-pulse-assisted readout, we perform resonator spectroscopy with the qubit prepared in $\ket{0}$ as a function of external flux bias, where we tune the flux bias by varying the flux pulse amplitude applied during the spectroscopy measurement. We compare this to the flux-dependent resonator spectroscopy measurements performed using a DC current bias to tune the qubit, see Fig.~\ref{fig:cal}(a). With a known periodicity, in aligning these curves we can translate the flux pulse amplitude uploaded to the waveform generator at room temperature to the effective value of $\Phi_{\mathrm{ext}}/\Phi_{0}$, as seen by the qubit. 

We additionally measure the assignment fidelity with a fixed integration time of 200 ns as a function of $\Phi_{\mathrm{ext}}/\Phi_{0}$ achieved through flux pulsing for various readout pulse amplitudes in order to choose the appropriate corresponding parameters. We observe the highest readout fidelities across all measured flux points when using a readout amplitude of $\varepsilon = 400$ mV, where this value corresponds to the amplitude of the waveform uploaded to the control hardware at room temperature. We subsequently convert this amplitude to an effective resonator photon occupation by quantifying the measurement efficiency of our setup.

We extract the measurement efficiency of our readout chain following the protocol presented in Ref.~\cite{bultink2018general}. We perform variable strength single-shot readout measurements from which we find the SNR according to Eq.~\ref{SNR} as a function of the variable measurement pulse amplitude, $\varepsilon$. We fit the data with a linear curve of the form SNR = $a\varepsilon$ and find $a = 35.47$, see Fig.~\ref{fig:cal}(b). We additionally perform a variable strength Ramsey measurement in which the qubit idles for a time $\tau_{\mathrm{idle}} = 1.52~\mu$s after the variable strength measurement pulse to allow for resonator photon depletion. The final $\pi/2$ rotation is performed about a variable axis of rotation defined by the azimuthal angle $\phi = [0, 4\pi]$ such that we measure $\langle\hat{\sigma}_{z}\rangle(\phi)$ for each value of $\varepsilon$. We fit the resulting oscillations with $A\mathrm{cos}(\phi+\phi_{0})+B$. From these fits, the qubit state's coherence is given by $\vert\rho_{01}\vert = A/2$, which we plot as a function of $\varepsilon$, see Fig.~\ref{fig:cal}(b). Fitting this data with a Gaussian of the form $\vert\rho_{01}(\varepsilon)\vert = \vert\rho_{01}(\varepsilon=0)\vert e^{-\varepsilon/(2\sigma^{2})}$, we find $\sigma = 6.93\times10^{-3}$.

The resulting measurement efficiency is given by $\eta = a^{2}\sigma^{2} = 6.04\%$. We can additionally use these results in order to translate the pulse amplitude, $\varepsilon$, defined by room temperature electronics to the effective resonator photon number, $\Bar{n}$. Assuming the dephasing rate, $\Gamma_{\mathrm{d}} = \frac{8\chi^{2}\Bar{n}}{\kappa}$, is governed by the population of residual measurement photons, we can estimate $\Bar{n}$ during the time between the two Ramsey pulses according to
\begin{equation}
    \Bar{n} = \frac{\varepsilon^{2}\kappa}{32\sigma^{2}\chi^{2}\tau_{\mathrm{total}}},
\end{equation}
\noindent where $\tau_{\mathrm{total}}$ is the duration of the measurement pulse (2.27 $\mu$s) summed with the idling time (1.52 $\mu$s). For an amplitude of $\varepsilon = 0.4$ V, we estimate $\Bar{n} = 31.2$ during $\tau_{\mathrm{total}}$. Further considering a square measurement pulse and a negligible resonator ring-up time ($5\kappa^{-1} \approx 130$ ns), this implies $\Bar{n} = 52.1$ when the measurement pulse is active.

Using the extracted measurement efficiency of 6.04$\%$ from this calibration in the numerical modeling of our results presented in Fig.~\ref{fig:readout}, we additionally offset the simulated curves by 40 ns to account for the time difference of the signal acquisition in the measurement efficiency calibration versus the single-shot readout measurements.

\section{Readout Parameter Overview}
\label{app:compare}
We summarize the qubit frequencies, dispersive shifts, and resonator linewidths of the fluxonium-resonator systems used to obtain the associated readout assignment fidelities reported in Refs.~\cite{bao2022fluxonium, ding2023high, kou2018simultaneous, moskalenko2022high, najera2024high, sun2023characterization} and visualized in Fig.~\ref{fig:compare}, as compared to this work, see Table~\ref{tab:lit readout}. We additionally include details pertaining to Refs.~\cite{wang2024efficient,zhang2021universal} and specify the type of parametric amplifier used in the readout chain of the corresponding setup where applicable.

\bibliography{bib.bib}

\begin{thebibliography}{56}%
\makeatletter
\providecommand \@ifxundefined [1]{%
 \@ifx{#1\undefined}
}%
\providecommand \@ifnum [1]{%
 \ifnum #1\expandafter \@firstoftwo
 \else \expandafter \@secondoftwo
 \fi
}%
\providecommand \@ifx [1]{%
 \ifx #1\expandafter \@firstoftwo
 \else \expandafter \@secondoftwo
 \fi
}%
\providecommand \natexlab [1]{#1}%
\providecommand \enquote  [1]{``#1''}%
\providecommand \bibnamefont  [1]{#1}%
\providecommand \bibfnamefont [1]{#1}%
\providecommand \citenamefont [1]{#1}%
\providecommand \href@noop [0]{\@secondoftwo}%
\providecommand \href [0]{\begingroup \@sanitize@url \@href}%
\providecommand \@href[1]{\@@startlink{#1}\@@href}%
\providecommand \@@href[1]{\endgroup#1\@@endlink}%
\providecommand \@sanitize@url [0]{\catcode `\\12\catcode `\$12\catcode `\&12\catcode `\#12\catcode `\^12\catcode `\_12\catcode `\%12\relax}%
\providecommand \@@startlink[1]{}%
\providecommand \@@endlink[0]{}%
\providecommand \url  [0]{\begingroup\@sanitize@url \@url }%
\providecommand \@url [1]{\endgroup\@href {#1}{\urlprefix }}%
\providecommand \urlprefix  [0]{URL }%
\providecommand \Eprint [0]{\href }%
\providecommand \doibase [0]{https://doi.org/}%
\providecommand \selectlanguage [0]{\@gobble}%
\providecommand \bibinfo  [0]{\@secondoftwo}%
\providecommand \bibfield  [0]{\@secondoftwo}%
\providecommand \translation [1]{[#1]}%
\providecommand \BibitemOpen [0]{}%
\providecommand \bibitemStop [0]{}%
\providecommand \bibitemNoStop [0]{.\EOS\space}%
\providecommand \EOS [0]{\spacefactor3000\relax}%
\providecommand \BibitemShut  [1]{\csname bibitem#1\endcsname}%
\let\auto@bib@innerbib\@empty
%</preamble>
\bibitem [{\citenamefont {Kjaergaard}\ \emph {et~al.}(2020)\citenamefont {Kjaergaard}, \citenamefont {Schwartz}, \citenamefont {Braum{\"u}ller}, \citenamefont {Krantz}, \citenamefont {Wang}, \citenamefont {Gustavsson},\ and\ \citenamefont {Oliver}}]{kjaergaard2020superconducting}%
  \BibitemOpen
  \bibfield  {author} {\bibinfo {author} {\bibfnamefont {M.}~\bibnamefont {Kjaergaard}}, \bibinfo {author} {\bibfnamefont {M.~E.}\ \bibnamefont {Schwartz}}, \bibinfo {author} {\bibfnamefont {J.}~\bibnamefont {Braum{\"u}ller}}, \bibinfo {author} {\bibfnamefont {P.}~\bibnamefont {Krantz}}, \bibinfo {author} {\bibfnamefont {J.~I.-J.}\ \bibnamefont {Wang}}, \bibinfo {author} {\bibfnamefont {S.}~\bibnamefont {Gustavsson}},\ and\ \bibinfo {author} {\bibfnamefont {W.~D.}\ \bibnamefont {Oliver}},\ }\bibfield  {title} {\bibinfo {title} {Superconducting qubits: Current state of play},\ }\href@noop {} {\bibfield  {journal} {\bibinfo  {journal} {Annual Review of Condensed Matter Physics}\ }\textbf {\bibinfo {volume} {11}},\ \bibinfo {pages} {369} (\bibinfo {year} {2020})}\BibitemShut {NoStop}%
\bibitem [{\citenamefont {Nakamura}\ \emph {et~al.}(1999)\citenamefont {Nakamura}, \citenamefont {Pashkin},\ and\ \citenamefont {Tsai}}]{nakamura1999coherent}%
  \BibitemOpen
  \bibfield  {author} {\bibinfo {author} {\bibfnamefont {Y.}~\bibnamefont {Nakamura}}, \bibinfo {author} {\bibfnamefont {Y.~A.}\ \bibnamefont {Pashkin}},\ and\ \bibinfo {author} {\bibfnamefont {J.}~\bibnamefont {Tsai}},\ }\bibfield  {title} {\bibinfo {title} {Coherent control of macroscopic quantum states in a single-cooper-pair box},\ }\href@noop {} {\bibfield  {journal} {\bibinfo  {journal} {nature}\ }\textbf {\bibinfo {volume} {398}},\ \bibinfo {pages} {786} (\bibinfo {year} {1999})}\BibitemShut {NoStop}%
\bibitem [{\citenamefont {Koch}\ \emph {et~al.}(2007)\citenamefont {Koch}, \citenamefont {Yu}, \citenamefont {Gambetta}, \citenamefont {Houck}, \citenamefont {Schuster}, \citenamefont {Majer}, \citenamefont {Blais}, \citenamefont {Devoret}, \citenamefont {Girvin},\ and\ \citenamefont {Schoelkopf}}]{koch2007charge}%
  \BibitemOpen
  \bibfield  {author} {\bibinfo {author} {\bibfnamefont {J.}~\bibnamefont {Koch}}, \bibinfo {author} {\bibfnamefont {T.~M.}\ \bibnamefont {Yu}}, \bibinfo {author} {\bibfnamefont {J.}~\bibnamefont {Gambetta}}, \bibinfo {author} {\bibfnamefont {A.~A.}\ \bibnamefont {Houck}}, \bibinfo {author} {\bibfnamefont {D.~I.}\ \bibnamefont {Schuster}}, \bibinfo {author} {\bibfnamefont {J.}~\bibnamefont {Majer}}, \bibinfo {author} {\bibfnamefont {A.}~\bibnamefont {Blais}}, \bibinfo {author} {\bibfnamefont {M.~H.}\ \bibnamefont {Devoret}}, \bibinfo {author} {\bibfnamefont {S.~M.}\ \bibnamefont {Girvin}},\ and\ \bibinfo {author} {\bibfnamefont {R.~J.}\ \bibnamefont {Schoelkopf}},\ }\bibfield  {title} {\bibinfo {title} {Charge-insensitive qubit design derived from the cooper pair box},\ }\href@noop {} {\bibfield  {journal} {\bibinfo  {journal} {Physical Review A—Atomic, Molecular, and Optical Physics}\ }\textbf {\bibinfo {volume} {76}},\ \bibinfo {pages} {042319} (\bibinfo {year} {2007})}\BibitemShut {NoStop}%
\bibitem [{\citenamefont {Manucharyan}\ \emph {et~al.}(2009)\citenamefont {Manucharyan}, \citenamefont {Koch}, \citenamefont {Glazman},\ and\ \citenamefont {Devoret}}]{manucharyan2009fluxonium}%
  \BibitemOpen
  \bibfield  {author} {\bibinfo {author} {\bibfnamefont {V.~E.}\ \bibnamefont {Manucharyan}}, \bibinfo {author} {\bibfnamefont {J.}~\bibnamefont {Koch}}, \bibinfo {author} {\bibfnamefont {L.~I.}\ \bibnamefont {Glazman}},\ and\ \bibinfo {author} {\bibfnamefont {M.~H.}\ \bibnamefont {Devoret}},\ }\bibfield  {title} {\bibinfo {title} {Fluxonium: Single cooper-pair circuit free of charge offsets},\ }\href@noop {} {\bibfield  {journal} {\bibinfo  {journal} {Science}\ }\textbf {\bibinfo {volume} {326}},\ \bibinfo {pages} {113} (\bibinfo {year} {2009})}\BibitemShut {NoStop}%
\bibitem [{\citenamefont {Martinis}\ \emph {et~al.}(2002)\citenamefont {Martinis}, \citenamefont {Nam}, \citenamefont {Aumentado},\ and\ \citenamefont {Urbina}}]{martinis2002rabi}%
  \BibitemOpen
  \bibfield  {author} {\bibinfo {author} {\bibfnamefont {J.~M.}\ \bibnamefont {Martinis}}, \bibinfo {author} {\bibfnamefont {S.}~\bibnamefont {Nam}}, \bibinfo {author} {\bibfnamefont {J.}~\bibnamefont {Aumentado}},\ and\ \bibinfo {author} {\bibfnamefont {C.}~\bibnamefont {Urbina}},\ }\bibfield  {title} {\bibinfo {title} {Rabi oscillations in a large josephson-junction qubit},\ }\href@noop {} {\bibfield  {journal} {\bibinfo  {journal} {Physical review letters}\ }\textbf {\bibinfo {volume} {89}},\ \bibinfo {pages} {117901} (\bibinfo {year} {2002})}\BibitemShut {NoStop}%
\bibitem [{\citenamefont {Orlando}\ \emph {et~al.}(1999)\citenamefont {Orlando}, \citenamefont {Mooij}, \citenamefont {Tian}, \citenamefont {Van Der~Wal}, \citenamefont {Levitov}, \citenamefont {Lloyd},\ and\ \citenamefont {Mazo}}]{orlando1999superconducting}%
  \BibitemOpen
  \bibfield  {author} {\bibinfo {author} {\bibfnamefont {T.}~\bibnamefont {Orlando}}, \bibinfo {author} {\bibfnamefont {J.}~\bibnamefont {Mooij}}, \bibinfo {author} {\bibfnamefont {L.}~\bibnamefont {Tian}}, \bibinfo {author} {\bibfnamefont {C.~H.}\ \bibnamefont {Van Der~Wal}}, \bibinfo {author} {\bibfnamefont {L.}~\bibnamefont {Levitov}}, \bibinfo {author} {\bibfnamefont {S.}~\bibnamefont {Lloyd}},\ and\ \bibinfo {author} {\bibfnamefont {J.}~\bibnamefont {Mazo}},\ }\bibfield  {title} {\bibinfo {title} {Superconducting persistent-current qubit},\ }\href@noop {} {\bibfield  {journal} {\bibinfo  {journal} {Physical Review B}\ }\textbf {\bibinfo {volume} {60}},\ \bibinfo {pages} {15398} (\bibinfo {year} {1999})}\BibitemShut {NoStop}%
\bibitem [{\citenamefont {Brooks}\ \emph {et~al.}(2013)\citenamefont {Brooks}, \citenamefont {Kitaev},\ and\ \citenamefont {Preskill}}]{brooks2013protected}%
  \BibitemOpen
  \bibfield  {author} {\bibinfo {author} {\bibfnamefont {P.}~\bibnamefont {Brooks}}, \bibinfo {author} {\bibfnamefont {A.}~\bibnamefont {Kitaev}},\ and\ \bibinfo {author} {\bibfnamefont {J.}~\bibnamefont {Preskill}},\ }\bibfield  {title} {\bibinfo {title} {Protected gates for superconducting qubits},\ }\href@noop {} {\bibfield  {journal} {\bibinfo  {journal} {Physical Review A—Atomic, Molecular, and Optical Physics}\ }\textbf {\bibinfo {volume} {87}},\ \bibinfo {pages} {052306} (\bibinfo {year} {2013})}\BibitemShut {NoStop}%
\bibitem [{\citenamefont {Krinner}\ \emph {et~al.}(2022)\citenamefont {Krinner}, \citenamefont {Lacroix}, \citenamefont {Remm}, \citenamefont {Di~Paolo}, \citenamefont {Genois}, \citenamefont {Leroux}, \citenamefont {Hellings}, \citenamefont {Lazar}, \citenamefont {Swiadek}, \citenamefont {Herrmann} \emph {et~al.}}]{krinner2022realizing}%
  \BibitemOpen
  \bibfield  {author} {\bibinfo {author} {\bibfnamefont {S.}~\bibnamefont {Krinner}}, \bibinfo {author} {\bibfnamefont {N.}~\bibnamefont {Lacroix}}, \bibinfo {author} {\bibfnamefont {A.}~\bibnamefont {Remm}}, \bibinfo {author} {\bibfnamefont {A.}~\bibnamefont {Di~Paolo}}, \bibinfo {author} {\bibfnamefont {E.}~\bibnamefont {Genois}}, \bibinfo {author} {\bibfnamefont {C.}~\bibnamefont {Leroux}}, \bibinfo {author} {\bibfnamefont {C.}~\bibnamefont {Hellings}}, \bibinfo {author} {\bibfnamefont {S.}~\bibnamefont {Lazar}}, \bibinfo {author} {\bibfnamefont {F.}~\bibnamefont {Swiadek}}, \bibinfo {author} {\bibfnamefont {J.}~\bibnamefont {Herrmann}}, \emph {et~al.},\ }\bibfield  {title} {\bibinfo {title} {Realizing repeated quantum error correction in a distance-three surface code},\ }\href@noop {} {\bibfield  {journal} {\bibinfo  {journal} {Nature}\ }\textbf {\bibinfo {volume} {605}},\ \bibinfo {pages} {669} (\bibinfo {year} {2022})}\BibitemShut {NoStop}%
\bibitem [{\citenamefont {Marques}\ \emph {et~al.}(2022)\citenamefont {Marques}, \citenamefont {Varbanov}, \citenamefont {Moreira}, \citenamefont {Ali}, \citenamefont {Muthusubramanian}, \citenamefont {Zachariadis}, \citenamefont {Battistel}, \citenamefont {Beekman}, \citenamefont {Haider}, \citenamefont {Vlothuizen} \emph {et~al.}}]{marques2022logical}%
  \BibitemOpen
  \bibfield  {author} {\bibinfo {author} {\bibfnamefont {J.~F.}\ \bibnamefont {Marques}}, \bibinfo {author} {\bibfnamefont {B.}~\bibnamefont {Varbanov}}, \bibinfo {author} {\bibfnamefont {M.}~\bibnamefont {Moreira}}, \bibinfo {author} {\bibfnamefont {H.}~\bibnamefont {Ali}}, \bibinfo {author} {\bibfnamefont {N.}~\bibnamefont {Muthusubramanian}}, \bibinfo {author} {\bibfnamefont {C.}~\bibnamefont {Zachariadis}}, \bibinfo {author} {\bibfnamefont {F.}~\bibnamefont {Battistel}}, \bibinfo {author} {\bibfnamefont {M.}~\bibnamefont {Beekman}}, \bibinfo {author} {\bibfnamefont {N.}~\bibnamefont {Haider}}, \bibinfo {author} {\bibfnamefont {W.}~\bibnamefont {Vlothuizen}}, \emph {et~al.},\ }\bibfield  {title} {\bibinfo {title} {Logical-qubit operations in an error-detecting surface code},\ }\href@noop {} {\bibfield  {journal} {\bibinfo  {journal} {Nature Physics}\ }\textbf {\bibinfo {volume} {18}},\ \bibinfo {pages} {80} (\bibinfo {year} {2022})}\BibitemShut {NoStop}%
\bibitem [{\citenamefont {Acharya}\ \emph {et~al.}(2024)\citenamefont {Acharya}, \citenamefont {Aghababaie-Beni}, \citenamefont {Aleiner}, \citenamefont {Andersen}, \citenamefont {Ansmann}, \citenamefont {Arute}, \citenamefont {Arya}, \citenamefont {Asfaw}, \citenamefont {Astrakhantsev}, \citenamefont {Atalaya} \emph {et~al.}}]{acharya2024quantum}%
  \BibitemOpen
  \bibfield  {author} {\bibinfo {author} {\bibfnamefont {R.}~\bibnamefont {Acharya}}, \bibinfo {author} {\bibfnamefont {L.}~\bibnamefont {Aghababaie-Beni}}, \bibinfo {author} {\bibfnamefont {I.}~\bibnamefont {Aleiner}}, \bibinfo {author} {\bibfnamefont {T.~I.}\ \bibnamefont {Andersen}}, \bibinfo {author} {\bibfnamefont {M.}~\bibnamefont {Ansmann}}, \bibinfo {author} {\bibfnamefont {F.}~\bibnamefont {Arute}}, \bibinfo {author} {\bibfnamefont {K.}~\bibnamefont {Arya}}, \bibinfo {author} {\bibfnamefont {A.}~\bibnamefont {Asfaw}}, \bibinfo {author} {\bibfnamefont {N.}~\bibnamefont {Astrakhantsev}}, \bibinfo {author} {\bibfnamefont {J.}~\bibnamefont {Atalaya}}, \emph {et~al.},\ }\bibfield  {title} {\bibinfo {title} {Quantum error correction below the surface code threshold},\ }\href@noop {} {\bibfield  {journal} {\bibinfo  {journal} {arXiv preprint arXiv:2408.13687}\ } (\bibinfo {year} {2024})}\BibitemShut {NoStop}%
\bibitem [{\citenamefont {Kim}\ \emph {et~al.}(2023)\citenamefont {Kim}, \citenamefont {Eddins}, \citenamefont {Anand}, \citenamefont {Wei}, \citenamefont {Van Den~Berg}, \citenamefont {Rosenblatt}, \citenamefont {Nayfeh}, \citenamefont {Wu}, \citenamefont {Zaletel}, \citenamefont {Temme} \emph {et~al.}}]{kim2023evidence}%
  \BibitemOpen
  \bibfield  {author} {\bibinfo {author} {\bibfnamefont {Y.}~\bibnamefont {Kim}}, \bibinfo {author} {\bibfnamefont {A.}~\bibnamefont {Eddins}}, \bibinfo {author} {\bibfnamefont {S.}~\bibnamefont {Anand}}, \bibinfo {author} {\bibfnamefont {K.~X.}\ \bibnamefont {Wei}}, \bibinfo {author} {\bibfnamefont {E.}~\bibnamefont {Van Den~Berg}}, \bibinfo {author} {\bibfnamefont {S.}~\bibnamefont {Rosenblatt}}, \bibinfo {author} {\bibfnamefont {H.}~\bibnamefont {Nayfeh}}, \bibinfo {author} {\bibfnamefont {Y.}~\bibnamefont {Wu}}, \bibinfo {author} {\bibfnamefont {M.}~\bibnamefont {Zaletel}}, \bibinfo {author} {\bibfnamefont {K.}~\bibnamefont {Temme}}, \emph {et~al.},\ }\bibfield  {title} {\bibinfo {title} {Evidence for the utility of quantum computing before fault tolerance},\ }\href@noop {} {\bibfield  {journal} {\bibinfo  {journal} {Nature}\ }\textbf {\bibinfo {volume} {618}},\ \bibinfo {pages} {500} (\bibinfo {year} {2023})}\BibitemShut {NoStop}%
\bibitem [{\citenamefont {Eun}\ \emph {et~al.}(2023)\citenamefont {Eun}, \citenamefont {Park}, \citenamefont {Seo}, \citenamefont {Choi},\ and\ \citenamefont {Hahn}}]{eun2023shape}%
  \BibitemOpen
  \bibfield  {author} {\bibinfo {author} {\bibfnamefont {S.}~\bibnamefont {Eun}}, \bibinfo {author} {\bibfnamefont {S.~H.}\ \bibnamefont {Park}}, \bibinfo {author} {\bibfnamefont {K.}~\bibnamefont {Seo}}, \bibinfo {author} {\bibfnamefont {K.}~\bibnamefont {Choi}},\ and\ \bibinfo {author} {\bibfnamefont {S.}~\bibnamefont {Hahn}},\ }\bibfield  {title} {\bibinfo {title} {Shape optimization of superconducting transmon qubits for low surface dielectric loss},\ }\href@noop {} {\bibfield  {journal} {\bibinfo  {journal} {Journal of Physics D: Applied Physics}\ }\textbf {\bibinfo {volume} {56}},\ \bibinfo {pages} {505306} (\bibinfo {year} {2023})}\BibitemShut {NoStop}%
\bibitem [{\citenamefont {Wang}\ \emph {et~al.}(2015)\citenamefont {Wang}, \citenamefont {Axline}, \citenamefont {Gao}, \citenamefont {Brecht}, \citenamefont {Chu}, \citenamefont {Frunzio}, \citenamefont {Devoret},\ and\ \citenamefont {Schoelkopf}}]{wang2015surface}%
  \BibitemOpen
  \bibfield  {author} {\bibinfo {author} {\bibfnamefont {C.}~\bibnamefont {Wang}}, \bibinfo {author} {\bibfnamefont {C.}~\bibnamefont {Axline}}, \bibinfo {author} {\bibfnamefont {Y.~Y.}\ \bibnamefont {Gao}}, \bibinfo {author} {\bibfnamefont {T.}~\bibnamefont {Brecht}}, \bibinfo {author} {\bibfnamefont {Y.}~\bibnamefont {Chu}}, \bibinfo {author} {\bibfnamefont {L.}~\bibnamefont {Frunzio}}, \bibinfo {author} {\bibfnamefont {M.}~\bibnamefont {Devoret}},\ and\ \bibinfo {author} {\bibfnamefont {R.~J.}\ \bibnamefont {Schoelkopf}},\ }\bibfield  {title} {\bibinfo {title} {Surface participation and dielectric loss in superconducting qubits},\ }\href@noop {} {\bibfield  {journal} {\bibinfo  {journal} {Applied Physics Letters}\ }\textbf {\bibinfo {volume} {107}} (\bibinfo {year} {2015})}\BibitemShut {NoStop}%
\bibitem [{\citenamefont {Nguyen}\ \emph {et~al.}(2019)\citenamefont {Nguyen}, \citenamefont {Lin}, \citenamefont {Somoroff}, \citenamefont {Mencia}, \citenamefont {Grabon},\ and\ \citenamefont {Manucharyan}}]{nguyen2019high}%
  \BibitemOpen
  \bibfield  {author} {\bibinfo {author} {\bibfnamefont {L.~B.}\ \bibnamefont {Nguyen}}, \bibinfo {author} {\bibfnamefont {Y.-H.}\ \bibnamefont {Lin}}, \bibinfo {author} {\bibfnamefont {A.}~\bibnamefont {Somoroff}}, \bibinfo {author} {\bibfnamefont {R.}~\bibnamefont {Mencia}}, \bibinfo {author} {\bibfnamefont {N.}~\bibnamefont {Grabon}},\ and\ \bibinfo {author} {\bibfnamefont {V.~E.}\ \bibnamefont {Manucharyan}},\ }\bibfield  {title} {\bibinfo {title} {High-coherence fluxonium qubit},\ }\href@noop {} {\bibfield  {journal} {\bibinfo  {journal} {Physical Review X}\ }\textbf {\bibinfo {volume} {9}},\ \bibinfo {pages} {041041} (\bibinfo {year} {2019})}\BibitemShut {NoStop}%
\bibitem [{\citenamefont {Ding}\ \emph {et~al.}(2023)\citenamefont {Ding}, \citenamefont {Hays}, \citenamefont {Sung}, \citenamefont {Kannan}, \citenamefont {An}, \citenamefont {Di~Paolo}, \citenamefont {Karamlou}, \citenamefont {Hazard}, \citenamefont {Azar}, \citenamefont {Kim} \emph {et~al.}}]{ding2023high}%
  \BibitemOpen
  \bibfield  {author} {\bibinfo {author} {\bibfnamefont {L.}~\bibnamefont {Ding}}, \bibinfo {author} {\bibfnamefont {M.}~\bibnamefont {Hays}}, \bibinfo {author} {\bibfnamefont {Y.}~\bibnamefont {Sung}}, \bibinfo {author} {\bibfnamefont {B.}~\bibnamefont {Kannan}}, \bibinfo {author} {\bibfnamefont {J.}~\bibnamefont {An}}, \bibinfo {author} {\bibfnamefont {A.}~\bibnamefont {Di~Paolo}}, \bibinfo {author} {\bibfnamefont {A.~H.}\ \bibnamefont {Karamlou}}, \bibinfo {author} {\bibfnamefont {T.~M.}\ \bibnamefont {Hazard}}, \bibinfo {author} {\bibfnamefont {K.}~\bibnamefont {Azar}}, \bibinfo {author} {\bibfnamefont {D.~K.}\ \bibnamefont {Kim}}, \emph {et~al.},\ }\bibfield  {title} {\bibinfo {title} {High-fidelity, frequency-flexible two-qubit fluxonium gates with a transmon coupler},\ }\href@noop {} {\bibfield  {journal} {\bibinfo  {journal} {Physical Review X}\ }\textbf {\bibinfo {volume} {13}},\ \bibinfo {pages} {031035} (\bibinfo {year} {2023})}\BibitemShut {NoStop}%
\bibitem [{\citenamefont {Somoroff}\ \emph {et~al.}(2023)\citenamefont {Somoroff}, \citenamefont {Ficheux}, \citenamefont {Mencia}, \citenamefont {Xiong}, \citenamefont {Kuzmin},\ and\ \citenamefont {Manucharyan}}]{somoroff2023millisecond}%
  \BibitemOpen
  \bibfield  {author} {\bibinfo {author} {\bibfnamefont {A.}~\bibnamefont {Somoroff}}, \bibinfo {author} {\bibfnamefont {Q.}~\bibnamefont {Ficheux}}, \bibinfo {author} {\bibfnamefont {R.~A.}\ \bibnamefont {Mencia}}, \bibinfo {author} {\bibfnamefont {H.}~\bibnamefont {Xiong}}, \bibinfo {author} {\bibfnamefont {R.}~\bibnamefont {Kuzmin}},\ and\ \bibinfo {author} {\bibfnamefont {V.~E.}\ \bibnamefont {Manucharyan}},\ }\bibfield  {title} {\bibinfo {title} {Millisecond coherence in a superconducting qubit},\ }\href@noop {} {\bibfield  {journal} {\bibinfo  {journal} {Physical Review Letters}\ }\textbf {\bibinfo {volume} {130}},\ \bibinfo {pages} {267001} (\bibinfo {year} {2023})}\BibitemShut {NoStop}%
\bibitem [{\citenamefont {Rower}\ \emph {et~al.}(2024)\citenamefont {Rower}, \citenamefont {Ding}, \citenamefont {Zhang}, \citenamefont {Hays}, \citenamefont {An}, \citenamefont {Harrington}, \citenamefont {Rosen}, \citenamefont {Gertler}, \citenamefont {Hazard}, \citenamefont {Niedzielski} \emph {et~al.}}]{rower2024suppressing}%
  \BibitemOpen
  \bibfield  {author} {\bibinfo {author} {\bibfnamefont {D.~A.}\ \bibnamefont {Rower}}, \bibinfo {author} {\bibfnamefont {L.}~\bibnamefont {Ding}}, \bibinfo {author} {\bibfnamefont {H.}~\bibnamefont {Zhang}}, \bibinfo {author} {\bibfnamefont {M.}~\bibnamefont {Hays}}, \bibinfo {author} {\bibfnamefont {J.}~\bibnamefont {An}}, \bibinfo {author} {\bibfnamefont {P.~M.}\ \bibnamefont {Harrington}}, \bibinfo {author} {\bibfnamefont {I.~T.}\ \bibnamefont {Rosen}}, \bibinfo {author} {\bibfnamefont {J.~M.}\ \bibnamefont {Gertler}}, \bibinfo {author} {\bibfnamefont {T.~M.}\ \bibnamefont {Hazard}}, \bibinfo {author} {\bibfnamefont {B.~M.}\ \bibnamefont {Niedzielski}}, \emph {et~al.},\ }\bibfield  {title} {\bibinfo {title} {Suppressing counter-rotating errors for fast single-qubit gates with fluxonium},\ }\href@noop {} {\bibfield  {journal} {\bibinfo  {journal} {arXiv preprint arXiv:2406.08295}\ } (\bibinfo {year} {2024})}\BibitemShut {NoStop}%
\bibitem [{\citenamefont {Moskalenko}\ \emph {et~al.}(2022)\citenamefont {Moskalenko}, \citenamefont {Simakov}, \citenamefont {Abramov}, \citenamefont {Grigorev}, \citenamefont {Moskalev}, \citenamefont {Pishchimova}, \citenamefont {Smirnov}, \citenamefont {Zikiy}, \citenamefont {Rodionov},\ and\ \citenamefont {Besedin}}]{moskalenko2022high}%
  \BibitemOpen
  \bibfield  {author} {\bibinfo {author} {\bibfnamefont {I.~N.}\ \bibnamefont {Moskalenko}}, \bibinfo {author} {\bibfnamefont {I.~A.}\ \bibnamefont {Simakov}}, \bibinfo {author} {\bibfnamefont {N.~N.}\ \bibnamefont {Abramov}}, \bibinfo {author} {\bibfnamefont {A.~A.}\ \bibnamefont {Grigorev}}, \bibinfo {author} {\bibfnamefont {D.~O.}\ \bibnamefont {Moskalev}}, \bibinfo {author} {\bibfnamefont {A.~A.}\ \bibnamefont {Pishchimova}}, \bibinfo {author} {\bibfnamefont {N.~S.}\ \bibnamefont {Smirnov}}, \bibinfo {author} {\bibfnamefont {E.~V.}\ \bibnamefont {Zikiy}}, \bibinfo {author} {\bibfnamefont {I.~A.}\ \bibnamefont {Rodionov}},\ and\ \bibinfo {author} {\bibfnamefont {I.~S.}\ \bibnamefont {Besedin}},\ }\bibfield  {title} {\bibinfo {title} {High fidelity two-qubit gates on fluxoniums using a tunable coupler},\ }\href@noop {} {\bibfield  {journal} {\bibinfo  {journal} {npj Quantum Information}\ }\textbf {\bibinfo {volume} {8}},\ \bibinfo {pages} {130} (\bibinfo {year} {2022})}\BibitemShut {NoStop}%
\bibitem [{\citenamefont {Dogan}\ \emph {et~al.}(2022)\citenamefont {Dogan}, \citenamefont {Rosenstock}, \citenamefont {Guevel}, \citenamefont {Xiong}, \citenamefont {Mencia}, \citenamefont {Somoroff}, \citenamefont {Nesterov}, \citenamefont {Vavilov}, \citenamefont {Manucharyan},\ and\ \citenamefont {Wang}}]{dogan2022demonstration}%
  \BibitemOpen
  \bibfield  {author} {\bibinfo {author} {\bibfnamefont {E.}~\bibnamefont {Dogan}}, \bibinfo {author} {\bibfnamefont {D.}~\bibnamefont {Rosenstock}}, \bibinfo {author} {\bibfnamefont {L.~L.}\ \bibnamefont {Guevel}}, \bibinfo {author} {\bibfnamefont {H.}~\bibnamefont {Xiong}}, \bibinfo {author} {\bibfnamefont {R.~A.}\ \bibnamefont {Mencia}}, \bibinfo {author} {\bibfnamefont {A.}~\bibnamefont {Somoroff}}, \bibinfo {author} {\bibfnamefont {K.~N.}\ \bibnamefont {Nesterov}}, \bibinfo {author} {\bibfnamefont {M.~G.}\ \bibnamefont {Vavilov}}, \bibinfo {author} {\bibfnamefont {V.~E.}\ \bibnamefont {Manucharyan}},\ and\ \bibinfo {author} {\bibfnamefont {C.}~\bibnamefont {Wang}},\ }\bibfield  {title} {\bibinfo {title} {Demonstration of the two-fluxonium cross-resonance gate},\ }\href@noop {} {\bibfield  {journal} {\bibinfo  {journal} {arXiv preprint arXiv:2204.11829}\ } (\bibinfo {year} {2022})}\BibitemShut {NoStop}%
\bibitem [{\citenamefont {Zhang}\ \emph {et~al.}(2021)\citenamefont {Zhang}, \citenamefont {Chakram}, \citenamefont {Roy}, \citenamefont {Earnest}, \citenamefont {Lu}, \citenamefont {Huang}, \citenamefont {Weiss}, \citenamefont {Koch},\ and\ \citenamefont {Schuster}}]{zhang2021universal}%
  \BibitemOpen
  \bibfield  {author} {\bibinfo {author} {\bibfnamefont {H.}~\bibnamefont {Zhang}}, \bibinfo {author} {\bibfnamefont {S.}~\bibnamefont {Chakram}}, \bibinfo {author} {\bibfnamefont {T.}~\bibnamefont {Roy}}, \bibinfo {author} {\bibfnamefont {N.}~\bibnamefont {Earnest}}, \bibinfo {author} {\bibfnamefont {Y.}~\bibnamefont {Lu}}, \bibinfo {author} {\bibfnamefont {Z.}~\bibnamefont {Huang}}, \bibinfo {author} {\bibfnamefont {D.}~\bibnamefont {Weiss}}, \bibinfo {author} {\bibfnamefont {J.}~\bibnamefont {Koch}},\ and\ \bibinfo {author} {\bibfnamefont {D.~I.}\ \bibnamefont {Schuster}},\ }\bibfield  {title} {\bibinfo {title} {Universal fast-flux control of a coherent, low-frequency qubit},\ }\href@noop {} {\bibfield  {journal} {\bibinfo  {journal} {Physical Review X}\ }\textbf {\bibinfo {volume} {11}},\ \bibinfo {pages} {011010} (\bibinfo {year} {2021})}\BibitemShut {NoStop}%
\bibitem [{\citenamefont {Gusenkova}\ \emph {et~al.}(2021)\citenamefont {Gusenkova}, \citenamefont {Spiecker}, \citenamefont {Gebauer}, \citenamefont {Willsch}, \citenamefont {Willsch}, \citenamefont {Valenti}, \citenamefont {Karcher}, \citenamefont {Gr{\"u}nhaupt}, \citenamefont {Takmakov}, \citenamefont {Winkel} \emph {et~al.}}]{gusenkova2021quantum}%
  \BibitemOpen
  \bibfield  {author} {\bibinfo {author} {\bibfnamefont {D.}~\bibnamefont {Gusenkova}}, \bibinfo {author} {\bibfnamefont {M.}~\bibnamefont {Spiecker}}, \bibinfo {author} {\bibfnamefont {R.}~\bibnamefont {Gebauer}}, \bibinfo {author} {\bibfnamefont {M.}~\bibnamefont {Willsch}}, \bibinfo {author} {\bibfnamefont {D.}~\bibnamefont {Willsch}}, \bibinfo {author} {\bibfnamefont {F.}~\bibnamefont {Valenti}}, \bibinfo {author} {\bibfnamefont {N.}~\bibnamefont {Karcher}}, \bibinfo {author} {\bibfnamefont {L.}~\bibnamefont {Gr{\"u}nhaupt}}, \bibinfo {author} {\bibfnamefont {I.}~\bibnamefont {Takmakov}}, \bibinfo {author} {\bibfnamefont {P.}~\bibnamefont {Winkel}}, \emph {et~al.},\ }\bibfield  {title} {\bibinfo {title} {Quantum nondemolition dispersive readout of a superconducting artificial atom using large photon numbers},\ }\href@noop {} {\bibfield  {journal} {\bibinfo  {journal} {Physical Review Applied}\ }\textbf {\bibinfo {volume} {15}},\ \bibinfo {pages} {064030} (\bibinfo {year} {2021})}\BibitemShut {NoStop}%
\bibitem [{\citenamefont {Gebauer}\ \emph {et~al.}(2020)\citenamefont {Gebauer}, \citenamefont {Karcher}, \citenamefont {Gusenkova}, \citenamefont {Spiecker}, \citenamefont {Gr{\"u}nhaupt}, \citenamefont {Takmakov}, \citenamefont {Winkel}, \citenamefont {Planat}, \citenamefont {Roch}, \citenamefont {Wernsdorfer} \emph {et~al.}}]{gebauer2020state}%
  \BibitemOpen
  \bibfield  {author} {\bibinfo {author} {\bibfnamefont {R.}~\bibnamefont {Gebauer}}, \bibinfo {author} {\bibfnamefont {N.}~\bibnamefont {Karcher}}, \bibinfo {author} {\bibfnamefont {D.}~\bibnamefont {Gusenkova}}, \bibinfo {author} {\bibfnamefont {M.}~\bibnamefont {Spiecker}}, \bibinfo {author} {\bibfnamefont {L.}~\bibnamefont {Gr{\"u}nhaupt}}, \bibinfo {author} {\bibfnamefont {I.}~\bibnamefont {Takmakov}}, \bibinfo {author} {\bibfnamefont {P.}~\bibnamefont {Winkel}}, \bibinfo {author} {\bibfnamefont {L.}~\bibnamefont {Planat}}, \bibinfo {author} {\bibfnamefont {N.}~\bibnamefont {Roch}}, \bibinfo {author} {\bibfnamefont {W.}~\bibnamefont {Wernsdorfer}}, \emph {et~al.},\ }\bibfield  {title} {\bibinfo {title} {State preparation of a fluxonium qubit with feedback from a custom fpga-based platform},\ }in\ \href@noop {} {\emph {\bibinfo {booktitle} {AIP Conference Proceedings}}},\ Vol.\ \bibinfo {volume} {2241}\ (\bibinfo {organization} {AIP Publishing},\ \bibinfo {year} {2020})\BibitemShut {NoStop}%
\bibitem [{\citenamefont {Wang}\ \emph {et~al.}(2024)\citenamefont {Wang}, \citenamefont {Wu}, \citenamefont {Wang}, \citenamefont {Ma}, \citenamefont {Zhang}, \citenamefont {Chen}, \citenamefont {Deng}, \citenamefont {Gao}, \citenamefont {Hu}, \citenamefont {Ma} \emph {et~al.}}]{wang2024efficient}%
  \BibitemOpen
  \bibfield  {author} {\bibinfo {author} {\bibfnamefont {T.}~\bibnamefont {Wang}}, \bibinfo {author} {\bibfnamefont {F.}~\bibnamefont {Wu}}, \bibinfo {author} {\bibfnamefont {F.}~\bibnamefont {Wang}}, \bibinfo {author} {\bibfnamefont {X.}~\bibnamefont {Ma}}, \bibinfo {author} {\bibfnamefont {G.}~\bibnamefont {Zhang}}, \bibinfo {author} {\bibfnamefont {J.}~\bibnamefont {Chen}}, \bibinfo {author} {\bibfnamefont {H.}~\bibnamefont {Deng}}, \bibinfo {author} {\bibfnamefont {R.}~\bibnamefont {Gao}}, \bibinfo {author} {\bibfnamefont {R.}~\bibnamefont {Hu}}, \bibinfo {author} {\bibfnamefont {L.}~\bibnamefont {Ma}}, \emph {et~al.},\ }\bibfield  {title} {\bibinfo {title} {Efficient initialization of fluxonium qubits based on auxiliary energy levels},\ }\href@noop {} {\bibfield  {journal} {\bibinfo  {journal} {Physical Review Letters}\ }\textbf {\bibinfo {volume} {132}},\ \bibinfo {pages} {230601} (\bibinfo {year} {2024})}\BibitemShut {NoStop}%
\bibitem [{\citenamefont {Stefanski}\ and\ \citenamefont {Andersen}(2024)}]{stefanski2024flux}%
  \BibitemOpen
  \bibfield  {author} {\bibinfo {author} {\bibfnamefont {T.~V.}\ \bibnamefont {Stefanski}}\ and\ \bibinfo {author} {\bibfnamefont {C.~K.}\ \bibnamefont {Andersen}},\ }\bibfield  {title} {\bibinfo {title} {Flux-pulse-assisted readout of a fluxonium qubit},\ }\href@noop {} {\bibfield  {journal} {\bibinfo  {journal} {Physical Review Applied}\ }\textbf {\bibinfo {volume} {22}},\ \bibinfo {pages} {014079} (\bibinfo {year} {2024})}\BibitemShut {NoStop}%
\bibitem [{\citenamefont {Swiadek}\ \emph {et~al.}(2023)\citenamefont {Swiadek}, \citenamefont {Shillito}, \citenamefont {Magnard}, \citenamefont {Remm}, \citenamefont {Hellings}, \citenamefont {Lacroix}, \citenamefont {Ficheux}, \citenamefont {Zanuz}, \citenamefont {Norris}, \citenamefont {Blais} \emph {et~al.}}]{swiadek2023enhancing}%
  \BibitemOpen
  \bibfield  {author} {\bibinfo {author} {\bibfnamefont {F.}~\bibnamefont {Swiadek}}, \bibinfo {author} {\bibfnamefont {R.}~\bibnamefont {Shillito}}, \bibinfo {author} {\bibfnamefont {P.}~\bibnamefont {Magnard}}, \bibinfo {author} {\bibfnamefont {A.}~\bibnamefont {Remm}}, \bibinfo {author} {\bibfnamefont {C.}~\bibnamefont {Hellings}}, \bibinfo {author} {\bibfnamefont {N.}~\bibnamefont {Lacroix}}, \bibinfo {author} {\bibfnamefont {Q.}~\bibnamefont {Ficheux}}, \bibinfo {author} {\bibfnamefont {D.~C.}\ \bibnamefont {Zanuz}}, \bibinfo {author} {\bibfnamefont {G.~J.}\ \bibnamefont {Norris}}, \bibinfo {author} {\bibfnamefont {A.}~\bibnamefont {Blais}}, \emph {et~al.},\ }\bibfield  {title} {\bibinfo {title} {Enhancing dispersive readout of superconducting qubits through dynamic control of the dispersive shift: Experiment and theory},\ }\href@noop {} {\bibfield  {journal} {\bibinfo  {journal} {arXiv preprint arXiv:2307.07765}\ } (\bibinfo {year} {2023})}\BibitemShut {NoStop}%
\bibitem [{\citenamefont {White}\ \emph {et~al.}(2023)\citenamefont {White}, \citenamefont {Opremcak}, \citenamefont {Sterling}, \citenamefont {Korotkov}, \citenamefont {Sank}, \citenamefont {Acharya}, \citenamefont {Ansmann}, \citenamefont {Arute}, \citenamefont {Arya}, \citenamefont {Bardin} \emph {et~al.}}]{white2023readout}%
  \BibitemOpen
  \bibfield  {author} {\bibinfo {author} {\bibfnamefont {T.}~\bibnamefont {White}}, \bibinfo {author} {\bibfnamefont {A.}~\bibnamefont {Opremcak}}, \bibinfo {author} {\bibfnamefont {G.}~\bibnamefont {Sterling}}, \bibinfo {author} {\bibfnamefont {A.}~\bibnamefont {Korotkov}}, \bibinfo {author} {\bibfnamefont {D.}~\bibnamefont {Sank}}, \bibinfo {author} {\bibfnamefont {R.}~\bibnamefont {Acharya}}, \bibinfo {author} {\bibfnamefont {M.}~\bibnamefont {Ansmann}}, \bibinfo {author} {\bibfnamefont {F.}~\bibnamefont {Arute}}, \bibinfo {author} {\bibfnamefont {K.}~\bibnamefont {Arya}}, \bibinfo {author} {\bibfnamefont {J.~C.}\ \bibnamefont {Bardin}}, \emph {et~al.},\ }\bibfield  {title} {\bibinfo {title} {Readout of a quantum processor with high dynamic range josephson parametric amplifiers},\ }\href@noop {} {\bibfield  {journal} {\bibinfo  {journal} {Applied Physics Letters}\ }\textbf {\bibinfo {volume} {122}} (\bibinfo {year} {2023})}\BibitemShut {NoStop}%
\bibitem [{\citenamefont {Macklin}\ \emph {et~al.}(2015)\citenamefont {Macklin}, \citenamefont {O’brien}, \citenamefont {Hover}, \citenamefont {Schwartz}, \citenamefont {Bolkhovsky}, \citenamefont {Zhang}, \citenamefont {Oliver},\ and\ \citenamefont {Siddiqi}}]{macklin2015near}%
  \BibitemOpen
  \bibfield  {author} {\bibinfo {author} {\bibfnamefont {C.}~\bibnamefont {Macklin}}, \bibinfo {author} {\bibfnamefont {K.}~\bibnamefont {O’brien}}, \bibinfo {author} {\bibfnamefont {D.}~\bibnamefont {Hover}}, \bibinfo {author} {\bibfnamefont {M.}~\bibnamefont {Schwartz}}, \bibinfo {author} {\bibfnamefont {V.}~\bibnamefont {Bolkhovsky}}, \bibinfo {author} {\bibfnamefont {X.}~\bibnamefont {Zhang}}, \bibinfo {author} {\bibfnamefont {W.}~\bibnamefont {Oliver}},\ and\ \bibinfo {author} {\bibfnamefont {I.}~\bibnamefont {Siddiqi}},\ }\bibfield  {title} {\bibinfo {title} {A near--quantum-limited josephson traveling-wave parametric amplifier},\ }\href@noop {} {\bibfield  {journal} {\bibinfo  {journal} {Science}\ }\textbf {\bibinfo {volume} {350}},\ \bibinfo {pages} {307} (\bibinfo {year} {2015})}\BibitemShut {NoStop}%
\bibitem [{\citenamefont {Frattini}\ \emph {et~al.}(2018)\citenamefont {Frattini}, \citenamefont {Sivak}, \citenamefont {Lingenfelter}, \citenamefont {Shankar},\ and\ \citenamefont {Devoret}}]{frattini2018optimizing}%
  \BibitemOpen
  \bibfield  {author} {\bibinfo {author} {\bibfnamefont {N.}~\bibnamefont {Frattini}}, \bibinfo {author} {\bibfnamefont {V.}~\bibnamefont {Sivak}}, \bibinfo {author} {\bibfnamefont {A.}~\bibnamefont {Lingenfelter}}, \bibinfo {author} {\bibfnamefont {S.}~\bibnamefont {Shankar}},\ and\ \bibinfo {author} {\bibfnamefont {M.}~\bibnamefont {Devoret}},\ }\bibfield  {title} {\bibinfo {title} {Optimizing the nonlinearity and dissipation of a snail parametric amplifier for dynamic range},\ }\href@noop {} {\bibfield  {journal} {\bibinfo  {journal} {Physical Review Applied}\ }\textbf {\bibinfo {volume} {10}},\ \bibinfo {pages} {054020} (\bibinfo {year} {2018})}\BibitemShut {NoStop}%
\bibitem [{\citenamefont {Vijay}\ \emph {et~al.}(2011)\citenamefont {Vijay}, \citenamefont {Slichter},\ and\ \citenamefont {Siddiqi}}]{vijay2011observation}%
  \BibitemOpen
  \bibfield  {author} {\bibinfo {author} {\bibfnamefont {R.}~\bibnamefont {Vijay}}, \bibinfo {author} {\bibfnamefont {D.}~\bibnamefont {Slichter}},\ and\ \bibinfo {author} {\bibfnamefont {I.}~\bibnamefont {Siddiqi}},\ }\bibfield  {title} {\bibinfo {title} {Observation of quantum jumps in a superconducting artificial atom},\ }\href@noop {} {\bibfield  {journal} {\bibinfo  {journal} {Physical review letters}\ }\textbf {\bibinfo {volume} {106}},\ \bibinfo {pages} {110502} (\bibinfo {year} {2011})}\BibitemShut {NoStop}%
\bibitem [{\citenamefont {Yamamoto}\ \emph {et~al.}(2008)\citenamefont {Yamamoto}, \citenamefont {Inomata}, \citenamefont {Watanabe}, \citenamefont {Matsuba}, \citenamefont {Miyazaki}, \citenamefont {Oliver}, \citenamefont {Nakamura},\ and\ \citenamefont {Tsai}}]{yamamoto2008flux}%
  \BibitemOpen
  \bibfield  {author} {\bibinfo {author} {\bibfnamefont {T.}~\bibnamefont {Yamamoto}}, \bibinfo {author} {\bibfnamefont {K.}~\bibnamefont {Inomata}}, \bibinfo {author} {\bibfnamefont {M.}~\bibnamefont {Watanabe}}, \bibinfo {author} {\bibfnamefont {K.}~\bibnamefont {Matsuba}}, \bibinfo {author} {\bibfnamefont {T.}~\bibnamefont {Miyazaki}}, \bibinfo {author} {\bibfnamefont {W.~D.}\ \bibnamefont {Oliver}}, \bibinfo {author} {\bibfnamefont {Y.}~\bibnamefont {Nakamura}},\ and\ \bibinfo {author} {\bibfnamefont {J.}~\bibnamefont {Tsai}},\ }\bibfield  {title} {\bibinfo {title} {Flux-driven josephson parametric amplifier},\ }\href@noop {} {\bibfield  {journal} {\bibinfo  {journal} {Applied Physics Letters}\ }\textbf {\bibinfo {volume} {93}} (\bibinfo {year} {2008})}\BibitemShut {NoStop}%
\bibitem [{\citenamefont {Abrams}\ \emph {et~al.}(2019)\citenamefont {Abrams}, \citenamefont {Didier}, \citenamefont {Caldwell}, \citenamefont {Johnson},\ and\ \citenamefont {Ryan}}]{abrams2019methods}%
  \BibitemOpen
  \bibfield  {author} {\bibinfo {author} {\bibfnamefont {D.~M.}\ \bibnamefont {Abrams}}, \bibinfo {author} {\bibfnamefont {N.}~\bibnamefont {Didier}}, \bibinfo {author} {\bibfnamefont {S.~A.}\ \bibnamefont {Caldwell}}, \bibinfo {author} {\bibfnamefont {B.~R.}\ \bibnamefont {Johnson}},\ and\ \bibinfo {author} {\bibfnamefont {C.~A.}\ \bibnamefont {Ryan}},\ }\bibfield  {title} {\bibinfo {title} {Methods for measuring magnetic flux crosstalk between tunable transmons},\ }\href@noop {} {\bibfield  {journal} {\bibinfo  {journal} {Physical Review Applied}\ }\textbf {\bibinfo {volume} {12}},\ \bibinfo {pages} {064022} (\bibinfo {year} {2019})}\BibitemShut {NoStop}%
\bibitem [{\citenamefont {Chitta}\ \emph {et~al.}(2022)\citenamefont {Chitta}, \citenamefont {Zhao}, \citenamefont {Huang}, \citenamefont {Mondragon-Shem},\ and\ \citenamefont {Koch}}]{chitta2022computer}%
  \BibitemOpen
  \bibfield  {author} {\bibinfo {author} {\bibfnamefont {S.~P.}\ \bibnamefont {Chitta}}, \bibinfo {author} {\bibfnamefont {T.}~\bibnamefont {Zhao}}, \bibinfo {author} {\bibfnamefont {Z.}~\bibnamefont {Huang}}, \bibinfo {author} {\bibfnamefont {I.}~\bibnamefont {Mondragon-Shem}},\ and\ \bibinfo {author} {\bibfnamefont {J.}~\bibnamefont {Koch}},\ }\bibfield  {title} {\bibinfo {title} {Computer-aided quantization and numerical analysis of superconducting circuits},\ }\href@noop {} {\bibfield  {journal} {\bibinfo  {journal} {New Journal of Physics}\ }\textbf {\bibinfo {volume} {24}},\ \bibinfo {pages} {103020} (\bibinfo {year} {2022})}\BibitemShut {NoStop}%
\bibitem [{\citenamefont {Groszkowski}\ and\ \citenamefont {Koch}(2021)}]{groszkowski2021scqubits}%
  \BibitemOpen
  \bibfield  {author} {\bibinfo {author} {\bibfnamefont {P.}~\bibnamefont {Groszkowski}}\ and\ \bibinfo {author} {\bibfnamefont {J.}~\bibnamefont {Koch}},\ }\bibfield  {title} {\bibinfo {title} {Scqubits: a python package for superconducting qubits},\ }\href@noop {} {\bibfield  {journal} {\bibinfo  {journal} {Quantum}\ }\textbf {\bibinfo {volume} {5}},\ \bibinfo {pages} {583} (\bibinfo {year} {2021})}\BibitemShut {NoStop}%
\bibitem [{\citenamefont {Zhu}\ \emph {et~al.}(2013)\citenamefont {Zhu}, \citenamefont {Ferguson}, \citenamefont {Manucharyan},\ and\ \citenamefont {Koch}}]{zhu2013circuit}%
  \BibitemOpen
  \bibfield  {author} {\bibinfo {author} {\bibfnamefont {G.}~\bibnamefont {Zhu}}, \bibinfo {author} {\bibfnamefont {D.~G.}\ \bibnamefont {Ferguson}}, \bibinfo {author} {\bibfnamefont {V.~E.}\ \bibnamefont {Manucharyan}},\ and\ \bibinfo {author} {\bibfnamefont {J.}~\bibnamefont {Koch}},\ }\bibfield  {title} {\bibinfo {title} {Circuit qed with fluxonium qubits: Theory of the dispersive regime},\ }\href@noop {} {\bibfield  {journal} {\bibinfo  {journal} {Physical Review B—Condensed Matter and Materials Physics}\ }\textbf {\bibinfo {volume} {87}},\ \bibinfo {pages} {024510} (\bibinfo {year} {2013})}\BibitemShut {NoStop}%
\bibitem [{\citenamefont {Rol}\ \emph {et~al.}(2019)\citenamefont {Rol}, \citenamefont {Battistel}, \citenamefont {Malinowski}, \citenamefont {Bultink}, \citenamefont {Tarasinski}, \citenamefont {Vollmer}, \citenamefont {Haider}, \citenamefont {Muthusubramanian}, \citenamefont {Bruno}, \citenamefont {Terhal} \emph {et~al.}}]{rol2019fast}%
  \BibitemOpen
  \bibfield  {author} {\bibinfo {author} {\bibfnamefont {M.}~\bibnamefont {Rol}}, \bibinfo {author} {\bibfnamefont {F.}~\bibnamefont {Battistel}}, \bibinfo {author} {\bibfnamefont {F.}~\bibnamefont {Malinowski}}, \bibinfo {author} {\bibfnamefont {C.}~\bibnamefont {Bultink}}, \bibinfo {author} {\bibfnamefont {B.}~\bibnamefont {Tarasinski}}, \bibinfo {author} {\bibfnamefont {R.}~\bibnamefont {Vollmer}}, \bibinfo {author} {\bibfnamefont {N.}~\bibnamefont {Haider}}, \bibinfo {author} {\bibfnamefont {N.}~\bibnamefont {Muthusubramanian}}, \bibinfo {author} {\bibfnamefont {A.}~\bibnamefont {Bruno}}, \bibinfo {author} {\bibfnamefont {B.}~\bibnamefont {Terhal}}, \emph {et~al.},\ }\bibfield  {title} {\bibinfo {title} {Fast, high-fidelity conditional-phase gate exploiting leakage interference in weakly anharmonic superconducting qubits},\ }\href@noop {} {\bibfield  {journal} {\bibinfo  {journal} {Physical review letters}\ }\textbf {\bibinfo {volume} {123}},\ \bibinfo {pages} {120502} (\bibinfo {year}
  {2019})}\BibitemShut {NoStop}%
\bibitem [{\citenamefont {Blais}\ \emph {et~al.}(2004)\citenamefont {Blais}, \citenamefont {Huang}, \citenamefont {Wallraff}, \citenamefont {Girvin},\ and\ \citenamefont {Schoelkopf}}]{blais2004cavity}%
  \BibitemOpen
  \bibfield  {author} {\bibinfo {author} {\bibfnamefont {A.}~\bibnamefont {Blais}}, \bibinfo {author} {\bibfnamefont {R.-S.}\ \bibnamefont {Huang}}, \bibinfo {author} {\bibfnamefont {A.}~\bibnamefont {Wallraff}}, \bibinfo {author} {\bibfnamefont {S.~M.}\ \bibnamefont {Girvin}},\ and\ \bibinfo {author} {\bibfnamefont {R.~J.}\ \bibnamefont {Schoelkopf}},\ }\bibfield  {title} {\bibinfo {title} {Cavity quantum electrodynamics for superconducting electrical circuits: An architecture for quantum computation},\ }\href@noop {} {\bibfield  {journal} {\bibinfo  {journal} {Physical Review A—Atomic, Molecular, and Optical Physics}\ }\textbf {\bibinfo {volume} {69}},\ \bibinfo {pages} {062320} (\bibinfo {year} {2004})}\BibitemShut {NoStop}%
\bibitem [{\citenamefont {Blais}\ \emph {et~al.}(2021)\citenamefont {Blais}, \citenamefont {Grimsmo}, \citenamefont {Girvin},\ and\ \citenamefont {Wallraff}}]{blais2021circuit}%
  \BibitemOpen
  \bibfield  {author} {\bibinfo {author} {\bibfnamefont {A.}~\bibnamefont {Blais}}, \bibinfo {author} {\bibfnamefont {A.~L.}\ \bibnamefont {Grimsmo}}, \bibinfo {author} {\bibfnamefont {S.~M.}\ \bibnamefont {Girvin}},\ and\ \bibinfo {author} {\bibfnamefont {A.}~\bibnamefont {Wallraff}},\ }\bibfield  {title} {\bibinfo {title} {Circuit quantum electrodynamics},\ }\href@noop {} {\bibfield  {journal} {\bibinfo  {journal} {Reviews of Modern Physics}\ }\textbf {\bibinfo {volume} {93}},\ \bibinfo {pages} {025005} (\bibinfo {year} {2021})}\BibitemShut {NoStop}%
\bibitem [{\citenamefont {Gambetta}\ \emph {et~al.}(2006)\citenamefont {Gambetta}, \citenamefont {Blais}, \citenamefont {Schuster}, \citenamefont {Wallraff}, \citenamefont {Frunzio}, \citenamefont {Majer}, \citenamefont {Devoret}, \citenamefont {Girvin},\ and\ \citenamefont {Schoelkopf}}]{gambetta2006qubit}%
  \BibitemOpen
  \bibfield  {author} {\bibinfo {author} {\bibfnamefont {J.}~\bibnamefont {Gambetta}}, \bibinfo {author} {\bibfnamefont {A.}~\bibnamefont {Blais}}, \bibinfo {author} {\bibfnamefont {D.~I.}\ \bibnamefont {Schuster}}, \bibinfo {author} {\bibfnamefont {A.}~\bibnamefont {Wallraff}}, \bibinfo {author} {\bibfnamefont {L.}~\bibnamefont {Frunzio}}, \bibinfo {author} {\bibfnamefont {J.}~\bibnamefont {Majer}}, \bibinfo {author} {\bibfnamefont {M.~H.}\ \bibnamefont {Devoret}}, \bibinfo {author} {\bibfnamefont {S.~M.}\ \bibnamefont {Girvin}},\ and\ \bibinfo {author} {\bibfnamefont {R.~J.}\ \bibnamefont {Schoelkopf}},\ }\bibfield  {title} {\bibinfo {title} {Qubit-photon interactions in a cavity: Measurement-induced dephasing and number splitting},\ }\href@noop {} {\bibfield  {journal} {\bibinfo  {journal} {Physical Review A—Atomic, Molecular, and Optical Physics}\ }\textbf {\bibinfo {volume} {74}},\ \bibinfo {pages} {042318} (\bibinfo {year} {2006})}\BibitemShut {NoStop}%
\bibitem [{\citenamefont {Khezri}\ \emph {et~al.}(2023)\citenamefont {Khezri}, \citenamefont {Opremcak}, \citenamefont {Chen}, \citenamefont {Miao}, \citenamefont {McEwen}, \citenamefont {Bengtsson}, \citenamefont {White}, \citenamefont {Naaman}, \citenamefont {Sank}, \citenamefont {Korotkov} \emph {et~al.}}]{khezri2023measurement}%
  \BibitemOpen
  \bibfield  {author} {\bibinfo {author} {\bibfnamefont {M.}~\bibnamefont {Khezri}}, \bibinfo {author} {\bibfnamefont {A.}~\bibnamefont {Opremcak}}, \bibinfo {author} {\bibfnamefont {Z.}~\bibnamefont {Chen}}, \bibinfo {author} {\bibfnamefont {K.~C.}\ \bibnamefont {Miao}}, \bibinfo {author} {\bibfnamefont {M.}~\bibnamefont {McEwen}}, \bibinfo {author} {\bibfnamefont {A.}~\bibnamefont {Bengtsson}}, \bibinfo {author} {\bibfnamefont {T.}~\bibnamefont {White}}, \bibinfo {author} {\bibfnamefont {O.}~\bibnamefont {Naaman}}, \bibinfo {author} {\bibfnamefont {D.}~\bibnamefont {Sank}}, \bibinfo {author} {\bibfnamefont {A.~N.}\ \bibnamefont {Korotkov}}, \emph {et~al.},\ }\bibfield  {title} {\bibinfo {title} {Measurement-induced state transitions in a superconducting qubit: Within the rotating-wave approximation},\ }\href@noop {} {\bibfield  {journal} {\bibinfo  {journal} {Physical Review Applied}\ }\textbf {\bibinfo {volume} {20}},\ \bibinfo {pages} {054008} (\bibinfo {year} {2023})}\BibitemShut {NoStop}%
\bibitem [{\citenamefont {Sank}\ \emph {et~al.}(2016)\citenamefont {Sank}, \citenamefont {Chen}, \citenamefont {Khezri}, \citenamefont {Kelly}, \citenamefont {Barends}, \citenamefont {Campbell}, \citenamefont {Chen}, \citenamefont {Chiaro}, \citenamefont {Dunsworth}, \citenamefont {Fowler} \emph {et~al.}}]{sank2016measurement}%
  \BibitemOpen
  \bibfield  {author} {\bibinfo {author} {\bibfnamefont {D.}~\bibnamefont {Sank}}, \bibinfo {author} {\bibfnamefont {Z.}~\bibnamefont {Chen}}, \bibinfo {author} {\bibfnamefont {M.}~\bibnamefont {Khezri}}, \bibinfo {author} {\bibfnamefont {J.}~\bibnamefont {Kelly}}, \bibinfo {author} {\bibfnamefont {R.}~\bibnamefont {Barends}}, \bibinfo {author} {\bibfnamefont {B.}~\bibnamefont {Campbell}}, \bibinfo {author} {\bibfnamefont {Y.}~\bibnamefont {Chen}}, \bibinfo {author} {\bibfnamefont {B.}~\bibnamefont {Chiaro}}, \bibinfo {author} {\bibfnamefont {A.}~\bibnamefont {Dunsworth}}, \bibinfo {author} {\bibfnamefont {A.}~\bibnamefont {Fowler}}, \emph {et~al.},\ }\bibfield  {title} {\bibinfo {title} {Measurement-induced state transitions in a superconducting qubit: Beyond the rotating wave approximation},\ }\href@noop {} {\bibfield  {journal} {\bibinfo  {journal} {Physical review letters}\ }\textbf {\bibinfo {volume} {117}},\ \bibinfo {pages} {190503} (\bibinfo {year} {2016})}\BibitemShut {NoStop}%
\bibitem [{\citenamefont {Nesterov}\ and\ \citenamefont {Pechenezhskiy}(2024)}]{nesterov2024measurement}%
  \BibitemOpen
  \bibfield  {author} {\bibinfo {author} {\bibfnamefont {K.~N.}\ \bibnamefont {Nesterov}}\ and\ \bibinfo {author} {\bibfnamefont {I.~V.}\ \bibnamefont {Pechenezhskiy}},\ }\bibfield  {title} {\bibinfo {title} {Measurement-induced state transitions in dispersive qubit readout schemes},\ }\href@noop {} {\bibfield  {journal} {\bibinfo  {journal} {arXiv preprint arXiv:2402.07360}\ } (\bibinfo {year} {2024})}\BibitemShut {NoStop}%
\bibitem [{\citenamefont {Didier}\ \emph {et~al.}(2015{\natexlab{a}})\citenamefont {Didier}, \citenamefont {Bourassa},\ and\ \citenamefont {Blais}}]{didier2015fast}%
  \BibitemOpen
  \bibfield  {author} {\bibinfo {author} {\bibfnamefont {N.}~\bibnamefont {Didier}}, \bibinfo {author} {\bibfnamefont {J.}~\bibnamefont {Bourassa}},\ and\ \bibinfo {author} {\bibfnamefont {A.}~\bibnamefont {Blais}},\ }\bibfield  {title} {\bibinfo {title} {Fast quantum nondemolition readout by parametric modulation of longitudinal qubit-oscillator interaction},\ }\href@noop {} {\bibfield  {journal} {\bibinfo  {journal} {Physical review letters}\ }\textbf {\bibinfo {volume} {115}},\ \bibinfo {pages} {203601} (\bibinfo {year} {2015}{\natexlab{a}})}\BibitemShut {NoStop}%
\bibitem [{\citenamefont {Didier}\ \emph {et~al.}(2015{\natexlab{b}})\citenamefont {Didier}, \citenamefont {Kamal}, \citenamefont {Oliver}, \citenamefont {Blais},\ and\ \citenamefont {Clerk}}]{didier2015heisenberg}%
  \BibitemOpen
  \bibfield  {author} {\bibinfo {author} {\bibfnamefont {N.}~\bibnamefont {Didier}}, \bibinfo {author} {\bibfnamefont {A.}~\bibnamefont {Kamal}}, \bibinfo {author} {\bibfnamefont {W.~D.}\ \bibnamefont {Oliver}}, \bibinfo {author} {\bibfnamefont {A.}~\bibnamefont {Blais}},\ and\ \bibinfo {author} {\bibfnamefont {A.~A.}\ \bibnamefont {Clerk}},\ }\bibfield  {title} {\bibinfo {title} {Heisenberg-limited qubit read-out with two-mode squeezed light},\ }\href@noop {} {\bibfield  {journal} {\bibinfo  {journal} {Physical review letters}\ }\textbf {\bibinfo {volume} {115}},\ \bibinfo {pages} {093604} (\bibinfo {year} {2015}{\natexlab{b}})}\BibitemShut {NoStop}%
\bibitem [{\citenamefont {Gardiner}\ and\ \citenamefont {Zoller}(2004)}]{gardiner2004quantum}%
  \BibitemOpen
  \bibfield  {author} {\bibinfo {author} {\bibfnamefont {C.}~\bibnamefont {Gardiner}}\ and\ \bibinfo {author} {\bibfnamefont {P.}~\bibnamefont {Zoller}},\ }\href@noop {} {\emph {\bibinfo {title} {Quantum noise: a handbook of Markovian and non-Markovian quantum stochastic methods with applications to quantum optics}}}\ (\bibinfo  {publisher} {Springer Science \& Business Media},\ \bibinfo {year} {2004})\BibitemShut {NoStop}%
\bibitem [{\citenamefont {Bao}\ \emph {et~al.}(2022)\citenamefont {Bao}, \citenamefont {Deng}, \citenamefont {Ding}, \citenamefont {Gao}, \citenamefont {Gao}, \citenamefont {Huang}, \citenamefont {Jiang}, \citenamefont {Ku}, \citenamefont {Li}, \citenamefont {Ma} \emph {et~al.}}]{bao2022fluxonium}%
  \BibitemOpen
  \bibfield  {author} {\bibinfo {author} {\bibfnamefont {F.}~\bibnamefont {Bao}}, \bibinfo {author} {\bibfnamefont {H.}~\bibnamefont {Deng}}, \bibinfo {author} {\bibfnamefont {D.}~\bibnamefont {Ding}}, \bibinfo {author} {\bibfnamefont {R.}~\bibnamefont {Gao}}, \bibinfo {author} {\bibfnamefont {X.}~\bibnamefont {Gao}}, \bibinfo {author} {\bibfnamefont {C.}~\bibnamefont {Huang}}, \bibinfo {author} {\bibfnamefont {X.}~\bibnamefont {Jiang}}, \bibinfo {author} {\bibfnamefont {H.-S.}\ \bibnamefont {Ku}}, \bibinfo {author} {\bibfnamefont {Z.}~\bibnamefont {Li}}, \bibinfo {author} {\bibfnamefont {X.}~\bibnamefont {Ma}}, \emph {et~al.},\ }\bibfield  {title} {\bibinfo {title} {Fluxonium: an alternative qubit platform for high-fidelity operations},\ }\href@noop {} {\bibfield  {journal} {\bibinfo  {journal} {Physical review letters}\ }\textbf {\bibinfo {volume} {129}},\ \bibinfo {pages} {010502} (\bibinfo {year} {2022})}\BibitemShut {NoStop}%
\bibitem [{\citenamefont {Kou}\ \emph {et~al.}(2018)\citenamefont {Kou}, \citenamefont {Smith}, \citenamefont {Vool}, \citenamefont {Pop}, \citenamefont {Sliwa}, \citenamefont {Hatridge}, \citenamefont {Frunzio},\ and\ \citenamefont {Devoret}}]{kou2018simultaneous}%
  \BibitemOpen
  \bibfield  {author} {\bibinfo {author} {\bibfnamefont {A.}~\bibnamefont {Kou}}, \bibinfo {author} {\bibfnamefont {W.}~\bibnamefont {Smith}}, \bibinfo {author} {\bibfnamefont {U.}~\bibnamefont {Vool}}, \bibinfo {author} {\bibfnamefont {I.}~\bibnamefont {Pop}}, \bibinfo {author} {\bibfnamefont {K.}~\bibnamefont {Sliwa}}, \bibinfo {author} {\bibfnamefont {M.}~\bibnamefont {Hatridge}}, \bibinfo {author} {\bibfnamefont {L.}~\bibnamefont {Frunzio}},\ and\ \bibinfo {author} {\bibfnamefont {M.}~\bibnamefont {Devoret}},\ }\bibfield  {title} {\bibinfo {title} {Simultaneous monitoring of fluxonium qubits in a waveguide},\ }\href@noop {} {\bibfield  {journal} {\bibinfo  {journal} {Physical Review Applied}\ }\textbf {\bibinfo {volume} {9}},\ \bibinfo {pages} {064022} (\bibinfo {year} {2018})}\BibitemShut {NoStop}%
\bibitem [{\citenamefont {Najera-Santos}\ \emph {et~al.}(2024)\citenamefont {Najera-Santos}, \citenamefont {Rousseau}, \citenamefont {Gerashchenko}, \citenamefont {Patange}, \citenamefont {Riva}, \citenamefont {Villiers}, \citenamefont {Briant}, \citenamefont {Cohadon}, \citenamefont {Heidmann}, \citenamefont {Palomo} \emph {et~al.}}]{najera2024high}%
  \BibitemOpen
  \bibfield  {author} {\bibinfo {author} {\bibfnamefont {B.-L.}\ \bibnamefont {Najera-Santos}}, \bibinfo {author} {\bibfnamefont {R.}~\bibnamefont {Rousseau}}, \bibinfo {author} {\bibfnamefont {K.}~\bibnamefont {Gerashchenko}}, \bibinfo {author} {\bibfnamefont {H.}~\bibnamefont {Patange}}, \bibinfo {author} {\bibfnamefont {A.}~\bibnamefont {Riva}}, \bibinfo {author} {\bibfnamefont {M.}~\bibnamefont {Villiers}}, \bibinfo {author} {\bibfnamefont {T.}~\bibnamefont {Briant}}, \bibinfo {author} {\bibfnamefont {P.-F.}\ \bibnamefont {Cohadon}}, \bibinfo {author} {\bibfnamefont {A.}~\bibnamefont {Heidmann}}, \bibinfo {author} {\bibfnamefont {J.}~\bibnamefont {Palomo}}, \emph {et~al.},\ }\bibfield  {title} {\bibinfo {title} {High-sensitivity ac-charge detection with a mhz-frequency fluxonium qubit},\ }\href@noop {} {\bibfield  {journal} {\bibinfo  {journal} {Physical Review X}\ }\textbf {\bibinfo {volume} {14}},\ \bibinfo {pages} {011007} (\bibinfo {year} {2024})}\BibitemShut {NoStop}%
\bibitem [{\citenamefont {Sun}\ \emph {et~al.}(2023)\citenamefont {Sun}, \citenamefont {Wu}, \citenamefont {Ku}, \citenamefont {Ma}, \citenamefont {Qin}, \citenamefont {Song}, \citenamefont {Wang}, \citenamefont {Zhang}, \citenamefont {Zhou}, \citenamefont {Shi} \emph {et~al.}}]{sun2023characterization}%
  \BibitemOpen
  \bibfield  {author} {\bibinfo {author} {\bibfnamefont {H.}~\bibnamefont {Sun}}, \bibinfo {author} {\bibfnamefont {F.}~\bibnamefont {Wu}}, \bibinfo {author} {\bibfnamefont {H.-S.}\ \bibnamefont {Ku}}, \bibinfo {author} {\bibfnamefont {X.}~\bibnamefont {Ma}}, \bibinfo {author} {\bibfnamefont {J.}~\bibnamefont {Qin}}, \bibinfo {author} {\bibfnamefont {Z.}~\bibnamefont {Song}}, \bibinfo {author} {\bibfnamefont {T.}~\bibnamefont {Wang}}, \bibinfo {author} {\bibfnamefont {G.}~\bibnamefont {Zhang}}, \bibinfo {author} {\bibfnamefont {J.}~\bibnamefont {Zhou}}, \bibinfo {author} {\bibfnamefont {Y.}~\bibnamefont {Shi}}, \emph {et~al.},\ }\bibfield  {title} {\bibinfo {title} {Characterization of loss mechanisms in a fluxonium qubit},\ }\href@noop {} {\bibfield  {journal} {\bibinfo  {journal} {Physical Review Applied}\ }\textbf {\bibinfo {volume} {20}},\ \bibinfo {pages} {034016} (\bibinfo {year} {2023})}\BibitemShut {NoStop}%
\bibitem [{\citenamefont {Eichler}\ and\ \citenamefont {Wallraff}(2014)}]{eichler2014controlling}%
  \BibitemOpen
  \bibfield  {author} {\bibinfo {author} {\bibfnamefont {C.}~\bibnamefont {Eichler}}\ and\ \bibinfo {author} {\bibfnamefont {A.}~\bibnamefont {Wallraff}},\ }\bibfield  {title} {\bibinfo {title} {Controlling the dynamic range of a josephson parametric amplifier},\ }\href@noop {} {\bibfield  {journal} {\bibinfo  {journal} {EPJ Quantum Technology}\ }\textbf {\bibinfo {volume} {1}},\ \bibinfo {pages} {1} (\bibinfo {year} {2014})}\BibitemShut {NoStop}%
\bibitem [{\citenamefont {Esposito}\ \emph {et~al.}(2021)\citenamefont {Esposito}, \citenamefont {Ranadive}, \citenamefont {Planat},\ and\ \citenamefont {Roch}}]{esposito2021perspective}%
  \BibitemOpen
  \bibfield  {author} {\bibinfo {author} {\bibfnamefont {M.}~\bibnamefont {Esposito}}, \bibinfo {author} {\bibfnamefont {A.}~\bibnamefont {Ranadive}}, \bibinfo {author} {\bibfnamefont {L.}~\bibnamefont {Planat}},\ and\ \bibinfo {author} {\bibfnamefont {N.}~\bibnamefont {Roch}},\ }\bibfield  {title} {\bibinfo {title} {Perspective on traveling wave microwave parametric amplifiers},\ }\href@noop {} {\bibfield  {journal} {\bibinfo  {journal} {Applied Physics Letters}\ }\textbf {\bibinfo {volume} {119}} (\bibinfo {year} {2021})}\BibitemShut {NoStop}%
\bibitem [{\citenamefont {Rol}\ \emph {et~al.}(2020)\citenamefont {Rol}, \citenamefont {Ciorciaro}, \citenamefont {Malinowski}, \citenamefont {Tarasinski}, \citenamefont {Sagastizabal}, \citenamefont {Bultink}, \citenamefont {Salathe}, \citenamefont {Haandb{\ae}k}, \citenamefont {Sedivy},\ and\ \citenamefont {DiCarlo}}]{rol2020time}%
  \BibitemOpen
  \bibfield  {author} {\bibinfo {author} {\bibfnamefont {M.~A.}\ \bibnamefont {Rol}}, \bibinfo {author} {\bibfnamefont {L.}~\bibnamefont {Ciorciaro}}, \bibinfo {author} {\bibfnamefont {F.~K.}\ \bibnamefont {Malinowski}}, \bibinfo {author} {\bibfnamefont {B.~M.}\ \bibnamefont {Tarasinski}}, \bibinfo {author} {\bibfnamefont {R.~E.}\ \bibnamefont {Sagastizabal}}, \bibinfo {author} {\bibfnamefont {C.~C.}\ \bibnamefont {Bultink}}, \bibinfo {author} {\bibfnamefont {Y.}~\bibnamefont {Salathe}}, \bibinfo {author} {\bibfnamefont {N.}~\bibnamefont {Haandb{\ae}k}}, \bibinfo {author} {\bibfnamefont {J.}~\bibnamefont {Sedivy}},\ and\ \bibinfo {author} {\bibfnamefont {L.}~\bibnamefont {DiCarlo}},\ }\bibfield  {title} {\bibinfo {title} {Time-domain characterization and correction of on-chip distortion of control pulses in a quantum processor},\ }\href@noop {} {\bibfield  {journal} {\bibinfo  {journal} {Applied Physics Letters}\ }\textbf {\bibinfo {volume} {116}} (\bibinfo {year} {2020})}\BibitemShut {NoStop}%
\bibitem [{dat()}]{dataset}%
  \BibitemOpen
  \href@noop {} {}\bibinfo {note} {Experimental data repository: \url{https://doi.org/10.4121/1092cb12-9198-4d43-8500-401c78a5dc15}}\BibitemShut {NoStop}%
\bibitem [{cod()}]{code_repo}%
  \BibitemOpen
  \href@noop {} {}\bibinfo {note} {Analysis code repository: \url{https://github.com/AndersenQubitLab/FPA-RO-experimental}}\BibitemShut {NoStop}%
\bibitem [{\citenamefont {{QuTech}}(2020{\natexlab{a}})}]{S4gDocumentation}%
  \BibitemOpen
  \bibfield  {author} {\bibinfo {author} {\bibnamefont {{QuTech}}},\ }\href@noop {} {\bibinfo {title} {S4g - 4 channel i source: Documentation for the spi rack}},\ \bibinfo {howpublished} {\url{https://qtwork.tudelft.nl/~mtiggelman/modules/i-source/s4g.html}} (\bibinfo {year} {2020}{\natexlab{a}}),\ \bibinfo {note} {accessed: 2024-09-20}\BibitemShut {NoStop}%
\bibitem [{\citenamefont {{QuTech}}(2020{\natexlab{b}})}]{SPIRackDocumentation}%
  \BibitemOpen
  \bibfield  {author} {\bibinfo {author} {\bibnamefont {{QuTech}}},\ }\href@noop {} {\bibinfo {title} {Spi rack: Documentation for the spi rack}},\ \bibinfo {howpublished} {\url{https://qtwork.tudelft.nl/~mtiggelman/spi-rack.html}} (\bibinfo {year} {2020}{\natexlab{b}}),\ \bibinfo {note} {accessed: 2024-09-20}\BibitemShut {NoStop}%
\bibitem [{\citenamefont {Bultink}\ \emph {et~al.}(2018)\citenamefont {Bultink}, \citenamefont {Tarasinski}, \citenamefont {Haandb{\ae}k}, \citenamefont {Poletto}, \citenamefont {Haider}, \citenamefont {Michalak}, \citenamefont {Bruno},\ and\ \citenamefont {DiCarlo}}]{bultink2018general}%
  \BibitemOpen
  \bibfield  {author} {\bibinfo {author} {\bibfnamefont {C.~C.}\ \bibnamefont {Bultink}}, \bibinfo {author} {\bibfnamefont {B.}~\bibnamefont {Tarasinski}}, \bibinfo {author} {\bibfnamefont {N.}~\bibnamefont {Haandb{\ae}k}}, \bibinfo {author} {\bibfnamefont {S.}~\bibnamefont {Poletto}}, \bibinfo {author} {\bibfnamefont {N.}~\bibnamefont {Haider}}, \bibinfo {author} {\bibfnamefont {D.}~\bibnamefont {Michalak}}, \bibinfo {author} {\bibfnamefont {A.}~\bibnamefont {Bruno}},\ and\ \bibinfo {author} {\bibfnamefont {L.}~\bibnamefont {DiCarlo}},\ }\bibfield  {title} {\bibinfo {title} {General method for extracting the quantum efficiency of dispersive qubit readout in circuit qed},\ }\href@noop {} {\bibfield  {journal} {\bibinfo  {journal} {Applied Physics Letters}\ }\textbf {\bibinfo {volume} {112}} (\bibinfo {year} {2018})}\BibitemShut {NoStop}%
\end{thebibliography}%

\end{document}